\documentclass[twocolumn,superscriptaddress,prl,aps,preprintnumbers,nofootinbib,longbibliography]{revtex4-2}
\usepackage{amsmath,amssymb,bm,graphicx,color,gensymb,bbold,appendix}
\usepackage[colorlinks=true, citecolor={blue!80!black}, urlcolor={blue!50!black}, linkcolor = {blue!80!black}]{hyperref}
\usepackage{comment}

\usepackage{pdfpages} % include pdfs
\usepackage{pgffor} % for loops

\makeatletter
\AtBeginDocument{\let\LS@rot\@undefined}
\makeatother

\pdfximage{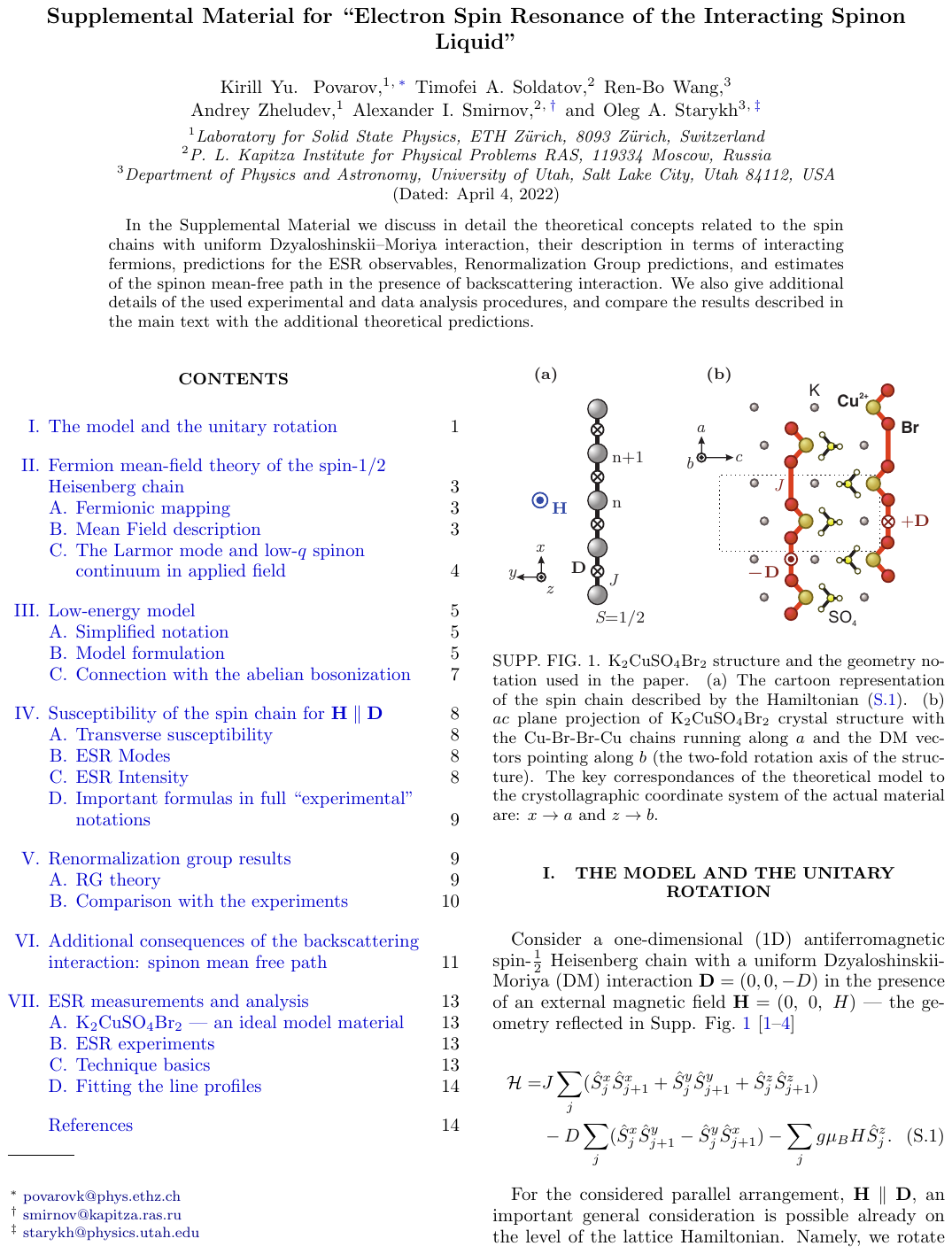}
\def\numbersupplementpages{\the\pdflastximagepages}

\newif\ifarXiv
\arXivtrue
\usepackage{xcolor}

\newcommand{\KCSOB}{K$_2$CuSO$_4$Br$_2$}

\newcommand{\hamilt}{\hat{\mathcal{H}}}

\newcommand{\spop}{\hat{\mathbf{S}}}

\bibpunct{[}{]}{,}{n}{}{}

\begin{document}
\title{Electron Spin Resonance of the Interacting Spinon Liquid}

\author{Kirill Yu. Povarov}
\email{povarovk@phys.ethz.ch}
\affiliation{Laboratory for Solid State Physics, ETH Z\"{u}rich, 8093 Z\"{u}rich, Switzerland}
%ORCID 0000-0002-0026-743X

\author{Timofei A. Soldatov}
\affiliation{P. L. Kapitza Institute for Physical Problems RAS, 119334 Moscow, Russia}

\author{Ren-Bo Wang}
\affiliation{Department of Physics and Astronomy, University of Utah, Salt Lake City, Utah 84112, USA} 

\author{Andrey Zheludev}
\affiliation{Laboratory for Solid State Physics, ETH Z\"{u}rich, 8093 Z\"{u}rich, Switzerland}
%\homepage{http://www.neutron.ethz.ch/}

\author{Alexander I. Smirnov}
 \email{smirnov@kapitza.ras.ru}
\affiliation{P. L. Kapitza Institute for Physical Problems RAS, 119334 Moscow, Russia} \email{smirnov@kapitza.ras.ru}

\author{Oleg A. Starykh}
\email{starykh@physics.utah.edu}
\affiliation{Department of Physics and Astronomy, University of Utah, Salt Lake City, Utah 84112, USA}

% ==============================================================================
\begin{abstract}
We report experimental verification of the recently predicted collective modes of spinons, stabilized by backscattering interaction, in a model quantum spin chain material. We
exploit the unique geometry of uniform Dzyaloshinskii--Moriya interactions in \KCSOB\ to measure the interaction-induced splitting between the two components of the electron spin resonance (ESR) response doublet. From that we
directly determine the magnitude of the ``marginally irrelevant'' backscattering interaction between spinons for the first time.
\end{abstract}

\date{\today}
\maketitle

Much of the current research in quantum magnetism is motivated by the search for an elusive quantum spin liquid (QSL) phase of the magnetic
matter. A salient feature of this entangled quantum state is the presence of fractionalized elementary excitations such as fermionic spin-1/2
spinons, interactions between which are mediated by the emergent gauge field \cite{Savary2017,ZhouKanodaNg_RMP_2017_SpinLiquidsReview}. This exotic, yet deeply rooted in history~\cite{Bethe1931,Faddeev1981,Pomeranchuk1941,Dzyaloshinskii1989,Anderson1973,*Anderson1987} perspective represents striking contrast with the
usual  integer-spin bosonic spin-wave excitations of the magnetically ordered media. It is firmly based on the remarkable experimental findings on (quasi) one-dimensional (1D) spin-$1/2$ magnetic insulators. These
include observations of a particle-hole continuum of excitations (also referred to as a ``two-spinon continuum") in the dynamical spin susceptibility $\chi(q,\omega)$ \cite{Tennant1995,Mourigal2013}  and magnetic field-controlled soft modes resulting from transitions on the Zeeman-split 1D Fermi surfaces
\cite{Dender1997,*Dender-thesis,*affleckoshikawa1999,Stone_PRL_2003_CupzNcontinua}.
The most recent milestone of this journey is provided by the Kitaev's honeycomb model which
harbors Majorana fermions as elementary excitations~\cite{Kitaev2006}. Experimental glimpses of this exciting physics~\cite{Wang2020,Kasahara2018,Bruin2021} continue to attract intense attention from the scientific community.

Close analogy between fractionalized excitations of the QSL and those of the standard Fermi liquid contains an important caveat. Unlike the
latter, elementary excitations of the QSL are highly nonlocal objects which appear and disappear only in pairs. Thus, the spinons cannot avoid interacting with each other. Interaction between spinons, as well as the curvature of the spinon
dispersion, determine shape of the continuum near its edges~\cite{Pereira2008,Imambekov2012}. In particular, one may expect a strong backscattering between the fermionic quasiparticles confined in 1D geometry. Still, being a ``marginally irrelevant'' interaction in the Renormalization Group (RG) sense~\cite{Eggert1996,Lukyanov1998}, the spinon backscattering manifests itself only through weak logarithmic corrections to the observables (e.g. the uniform spin susceptibility~\cite{Eggert1994,*Motoyama1996}, or the nuclear magnetic relaxation rate~\cite{Takigawa1996,*Takigawa1997,*Barzykin2001}) and is barely detectable this way. As the very recent theoretical findings show, it becomes most important when magnetic
field is applied, shifting spinon continuum up in energy and
producing a spin-1 oscillatory collective mode of spinons~\cite{Keselman2020} that originates from the Larmor frequency,
a spin chain analog of the Silin spin wave in nonferromagnetic metals~\cite{Silin1958,Silin1959,Platzman1967,Schultz1967,Leggett1970}. The backscattering interaction is straightforwardly manifest here through qualitative spectrum modifications~\cite{Keselman2020,WangKeselmanStarykh_arXiv_2022_DMhydrodynamics}.
However, direct observation of this novel effect with e.g. neutrons is a challenging task
that requires thoroughly balancing the field strength, the magnetic energy scale of the material, and the instrument resolution. Yet, alternative spectroscopic methods can overcome these difficulties.

In this Letter, for the first time, we experimentally investigate this interaction-induced modification of the spinon continuum with the help of
the electron spin resonance (ESR) technique. Our measurements lead to the direct and unambiguous determination of the backscattering interaction between
fractionalized spinon excitations of the spin chain. This finding is facilitated by the unique feature of the material --- the uniform
Dzyaloshinskii--Moriya (DM) interaction~\cite{Dzyaloshinskii_JChemPhysSol_1958_DM,Moriya_PR_1960_DM}.

\begin{figure*}
\centering
    \includegraphics[width=0.99\textwidth]{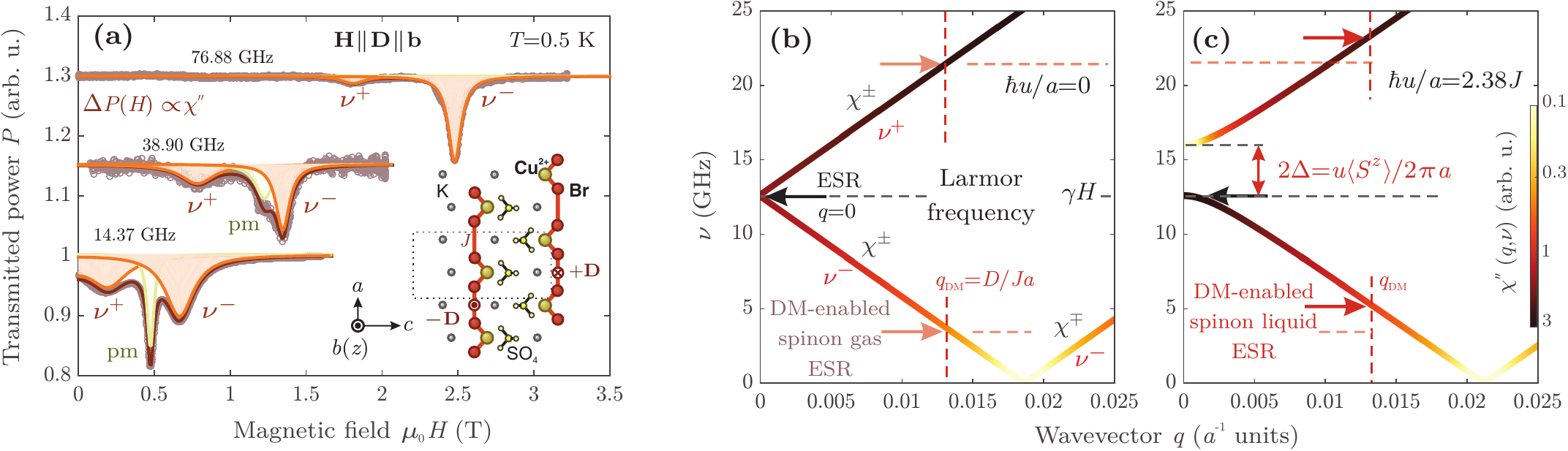}
    \caption{Electron spin resonance in \KCSOB.
    (a) Several low-$T$ resonance lines. Magnetic field is applied along the DM axis $b$. Gray points is the measured rate of absorption of the microwave radiation by the sample. It is well fitted by several Lorentzian lines (dark red line). The contributions of modes $\nu^+$ and $\nu^-$ is shown in orange color. At low frequencies an additional parasitic paramagnetic resonance can be detected (light green color). The inset shows the sketch of \KCSOB\ crystal structure.
    (b) Calculated spectrum of small-$q$ transverse spin fluctuations in a magnetized spin chain without the backscattering interaction. (c) The same for the interacting spinons,  $\hbar u=2.38 J a$. For \KCSOB\ considered here $J=20.5$~K, and the magnetic field is $0.3$~T in both panels. The solid lines show $\nu(q)$ for the poles of the transverse spin susceptibility according to Refs.~\cite{Keselman2020,SM}, and the color shows their intensity. Black (red) horizontal arrows indicate ESR frequencies of the spin chain without (black) and with (red)  the DM interaction.
      }
    \label{FIG:lines}
\end{figure*}

The material of our study is \KCSOB, providing
an outstanding realization of the $S=1/2$ Heisenberg chain antiferromagnet perturbed by a small uniform DM interaction~\cite{HaelgLorenz_PRB_2014_KCSOX,Haelg_2015_PhDthesis,Giovazzo_ZKrist_1976_KCSOC,JinStarykh_PRB_2017_KCSOXdiagram}. The magnetic Cu$^{2+}$ spin-$1/2$ ions at distance $a=7.73$~\AA\ to each other are forming linear chains running
along the $\mathbf{a}$ axis of the crystal [see inset of Fig.~\ref{FIG:lines}(a)].
Antiferromagnetic interaction $J\simeq20.5$~K~\cite{HaelgLorenz_PRB_2014_KCSOX} is mediated by a two-bromine unit, which lacks an inversion center
within the $ac$ plane. This lack of inversion symmetry naturally gives rise to the small DM interaction directed along the $\mathbf{b}$ axis. Hence, it can be described as a Heisenberg spin-1/2 chain, with exchange interaction $J$ between nearest-neighbor spins,  perturbed by the uniform DM interaction ${\bf D} \cdot \spop_{n}\times\spop_{n+1}$ and subject to the external magnetic field $\mathbf{H}$. We focus on the parallel geometry when the magnetic field is aligned along the DM axis ($z$-axis), $\mathbf{H} \parallel \mathbf{D} \parallel \hat{z}$ , which preserves the symmetry of rotations about $z$. The Hamiltonian reads

\begin{equation}
\label{EQ:AFplusDM}
\hamilt=\sum\limits_{n} J \spop_{n}\cdot\spop_{n+1} - D \hat{z} \cdot \spop_{n}\times\spop_{n+1} -g\mu_{B} H \hat{S}^z_n .
%D [\spop_{n}\times\spop_{n+1}]^z
\end{equation}

Semiclassically, the competition between the antiferromagnetic Heisenberg exchange $J$ and DM interaction
$\mathbf{D}$ results in an incommensurate spiral, the period of which is determined by the wave vector
$q_{\mathrm{DM}} = \tan^{-1}(D/J)/a \approx D/(J a)$. Quantum mechanically one can employ the
unitary position-dependent rotation of spins $\hat{S}^+_n = \hat{\tilde{S}}^+_n e^{-i q_{\rm DM} n a}$ which eliminates the DM term from the Hamiltonian
\eqref{EQ:AFplusDM} for the price of the momentum boost $q \to q + q_{\rm DM}$~\cite{GangadharaiahSunStarykh_PRB_2008_chainDM}~\cite{SM}.

A \emph{uniform}, bond-independent arrangement of DM vectors within the chain is a very rare occasion. However, a truly
remarkable property of \KCSOB\ that distinguishes it from similar materials (e.g.
Cs$_2$CuCl$_4$~\cite{Starykh_PRB_2010_Cs2CuCl4theory,PovarovSmirnov_PRL_2011_ESRdoublet}) is that the DM axis is the {\em same}, i.e. oriented
along the $b$ crystal axis, for all spin chains \cite{HaelgLorenz_PRB_2014_KCSOX}. This unique feature allows us to realize $\mathbf{H}\parallel\mathbf{D}$ geometry experimentally, which is a crucial element of our study. In this case the energy absorption rate measured by ESR is in fact determined by
${\rm Im}[\tilde{\chi}^\pm(q_{\mathrm{DM}},\nu)]$  \cite{GangadharaiahSunStarykh_PRB_2008_chainDM,Karimi2011}
--- the ESR becomes a finite momentum probe of the dynamic spin susceptibility!

The response of the chain~(\ref{EQ:AFplusDM}) at small momenta can in turn be understood in terms of fermion quasiparticles - spinons~\cite{Arovas1988,Keselman2020}. In the low-energy continuum limit the Heisenberg spin-1/2 chain is described by the field theory of
two component Dirac spinors $\hat{\psi}_{R/L} = (\hat{\psi}_{R/L, \uparrow}, \hat{\psi}_{R/L,\downarrow})^{\rm T}$~\cite{gogolin-book,*Garate2010,*Chan2017}.
Operators $\hat{\psi}_{R/L, s}$ describe spin-up ($s=\uparrow$) and spin-down ($s=\downarrow$) fermions with wave vectors near the right and left Fermi points $\pm k_{F}$ of the 1D Fermi surface. Uniform spin fluctuations are represented by the
spin current operators $\hat{{\bf J}}_{R/L} = \frac{1}{2} \hat{\psi}_{R/L}^\dagger {\bm \sigma} \hat{\psi}_{R/L}$. The Hamiltonian is written as the sum of two terms,
$\hamilt = \hamilt_0 + \hat{V}_{\rm bs}$,
\begin{eqnarray}
\label{eq:H0}
\hamilt_0 &=& \int dx \, \Big[\hbar v_F\Big(\hat{\psi}_R^\dagger(x) (-i \partial_x) \hat{\psi}_R(x) + \hat{\psi}_L^\dagger(x) (i \partial_x) \hat{\psi}_L(x) \Big)\nonumber\\
&& - g\mu_B H \left(\hat{J}^z_R(x) + \hat{J}^z_L(x)\right)\Big], \\
\hat{V}_{\rm bs} &=& - \hbar u \int dx \, \hat{{\bf J}}_{R}(x) \cdot \hat{{\bf J}}_{L}(x). 
\label{eq:Hbs}
\end{eqnarray}
Here $v_F = \pi J a/(2\hbar)$ is the spinon Fermi velocity, and $u$ denotes the backscattering (also known as the current-current) interaction between spinons. Despite its somewhat complicated appearance $\hamilt_0$ describes a non-interacting gas of neutral fermions (spinons) $\psi_{R/L, s}$ with linear dispersion, subject to
the external magnetic field. Interaction between spinons is compactly encoded in $\hat{V}_{\rm bs}$, which describes $2k_F$ scattering of right- and left-moving fermions
on each other. The amplitude $u$ of this backscattering \emph{is the key parameter that we experimentally address in our study}.

\begin{figure*}
\centering
    \includegraphics[width=0.99\textwidth]{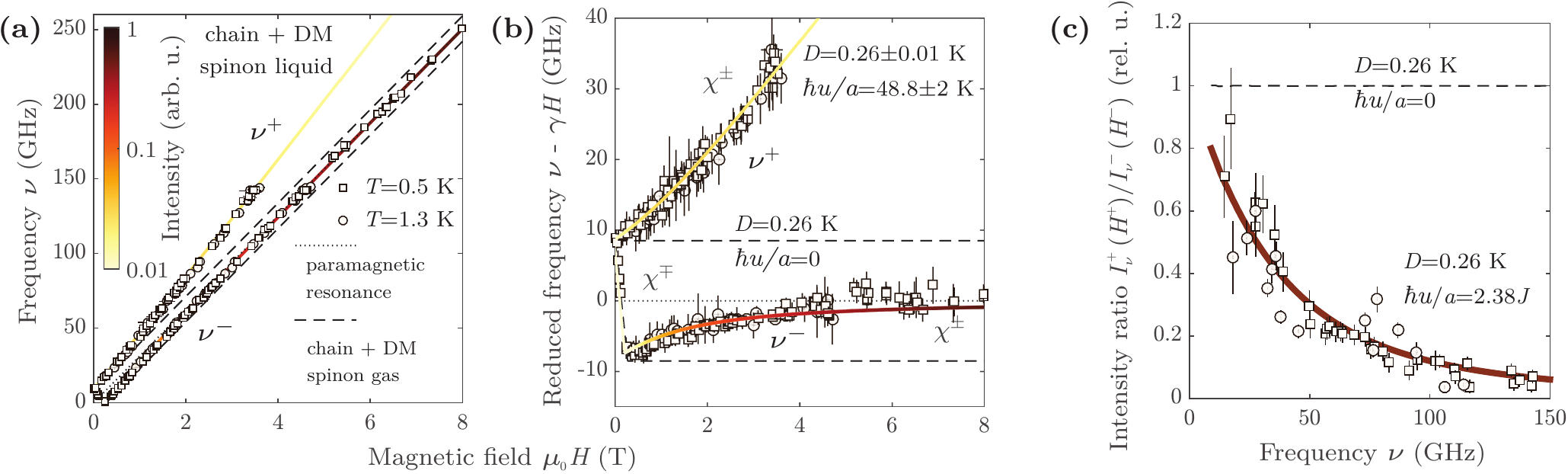}
        \caption{ESR spectrum of \KCSOB\ and its theoretical description in terms of interacting spinons.
        (a) The frequency-field diagram including the $H\parallel D$ data at two different temperatures (circles and squares). Error bars are obtained from the lineshape fit and for the majority of ESR lines are estimated as 0.02 T and are within the
symbol size. For some of frequencies above $50$ GHz, the line shape was distorted by the parasitic electrodynamic size effect, here
errors in $H$-value are estimated as  a whole linewidth and are seen as error bars on the left and middle panels.
        The  dotted line is the $h\nu=g \mu_B H$ paramagnetic resonance, the dashed lines correspond to the non-interacting approximation
         \eqref{EQ:simpleDoublet} with $D=0.26$~K. The solid lines and their color (intensity, as indicated in the inset)
          correspond to the interacting spinon theory \eqref{EQ:complexDoublet}. The best-fit value of $\hbar u/a=48.8$~K is used.
        (b) The same diagram, but in the ``reduced'' representation with $\gamma H$ subtracted, $\nu \to \nu - \gamma H$. (c) The mode intensity ratio
        versus the observation frequency. Points are the experimental data (as in the other panels), error bars stem from the experimental uncertainty in determining the widths and amplitudes of the two overlapping peaks. Solid line is the theoretical prediction \eqref{EQ:intensity2}
         for the obtained best-fit value $\delta=0.12$. The dashed line shows noninteracting spinons result. }
         \label{FIG:constantfits}
\end{figure*}

In the absence of the backscattering $u$ interaction, the ESR spectrum of the spin chain~\eqref{EQ:AFplusDM} is known to be a {\em
doublet}~\cite{GangadharaiahSunStarykh_PRB_2008_chainDM}
\begin{equation}
\label{EQ:simpleDoublet} h\nu^{\pm}= |g\mu_{B}H \pm \frac{\pi}{2}D|,
\end{equation}
accounting for both transverse $\chi^{\pm}$ and $\chi^{\mp}$ components of the susceptibility. Its origin can be easily understood from the low-energy spectrum of the spin chain depicted in Fig.~\ref{FIG:lines}(b). The appearance of the ESR doublet in the quantum spin liquid state of the spin chain is the fingerprint of the uniform DM interaction. It has been
experimentally observed previously in three different materials, Cs$_2$CuCl$_4$~\cite{PovarovSmirnov_PRL_2011_ESRdoublet}, the present compound
\KCSOB\ \cite{Smirnov2015}, and also in K$_2$CuSO$_4$Cl$_2$~\cite{Soldatov2018}.

However, the noninteracting spinon description of the dynamic spin response is qualitatively incomplete. Similar to the case of the interacting electron
liquid \cite{Silin1958,Silin1959}, the backscattering interaction $u$ qualitatively changes transverse spin susceptibility
for small $(q,\nu)$, as demonstrated by the recent interacting spinon theory \cite{Keselman2020}, supported by DMRG calculations and numerical Bethe-ansatz study \cite{Kohno2009}. The new poles of $\chi^{\pm, \mp}$ are shown in Fig.~\ref{FIG:lines}(c). The ESR frequencies are now given by \cite{Keselman2020}
\begin{equation}
\label{EQ:complexDoublet} h\nu^{\pm}=\left|g\mu_{B}H+\Delta\pm\sqrt{\Delta^{2}+(1-\delta^2) \Big(\frac{\pi}{2}D\Big)^2}\right|.
\end{equation}
Here $2\Delta=\hbar u \langle \hat{S}^{z}\rangle/a$ 
is the interaction-induced spectral gap (splitting), and $\langle \hat{S}^{z}\rangle/a = \chi H$ is the mean spin $z-$component per unit length. The theory is most compactly expressed in terms of the small dimensionless parameter $\delta$,
\begin{equation}
\label{EQ:deltadef0}
    \delta=\frac{1}{2a} \hbar u \frac{\chi_0}{g\mu_{B}} =\frac{u}{4\pi v_{F}} ,
\end{equation}
that describes the enhancement of the renormalized zero-field spin susceptibility per unit length $\chi = \chi_0/(1-\delta)$, from its
noninteracting value $\chi_0=g\mu_{B}a/(2\pi \hbar v_{F})$. Using $\delta$, the splitting becomes:
\begin{equation}
\label{EQ:deltadef}
    \Delta=\frac{\delta}{1-\delta}g\mu_{B}H.
\end{equation}
Even when $D=0$, Eq.\eqref{EQ:complexDoublet} predicts the finite spectral gap $2\Delta$ between the $\nu^{\pm}$ branches, as Fig.~\ref{FIG:lines}(c)
shows. 

This feature does not contradict the Larmor
theorem because the intensity of the upper branch $\nu^{+}$ vanishes as $q^2$ in the $q\to 0$ limit,
 whereas the lower intense branch $\nu^-$remains exactly at the Larmor frequency $h^{-1}g\mu_{B}H=\gamma H$. Hence, in the absence of the symmetry-breaking
 perturbations the Larmor theorem is actually obeyed, and backscattering interaction $u$ makes no difference for the ESR experiment on the {\em ideal} spin chain. Thus, the symmetry-breaking uniform DM interaction in Eq.~\eqref{EQ:AFplusDM} is absolutely crucial for accessing both modes with ESR.

guz5dune
The experiments were done at the Kapitza Institute  on a set of multifrequency (1-250 GHz) resonant cavity ESR inserts into $^3$He and $^4$He-pumping
cryostats equipped with superconducting magnets. The transmission of
microwave power $P$ through the sample-containing resonator was measured as the function of the magnetic field at a fixed frequency $\nu$.
It is affected by the dissipative susceptibility of the spin
subsystem and can be approximately expressed as $\Delta P/P\propto \chi''(0,\nu)$~\cite{Poole_1997_ESRbook}, or rather $\chi''(q_\mathrm{DM},\nu)$ in the presence of the
uniform DM interaction. To compare these predictions with the ESR experiments on  \KCSOB\, we use two sets of data. One, taken at $T=0.5$~K, was previously partly
described in Ref.~\cite{Smirnov2015}. The other set, taken at at $T=1.3$~K, was not presented before. The datasets involve multiple samples of \KCSOB.

Several examples of the raw spectrometer microwave transmission  data at $0.5$~K are shown in Fig.~\ref{FIG:lines}(a). The
data demonstrate a well-resolved doublet of $\nu^{\pm}$ lines, with a parasitic line (not exceeding $25$\% of $\nu$-lines intensity) in the middle that comes from impurities and is
sample dependent. The intensity of this mode shows Curie-like temperature dependence typical of impurity spins (see  Ref.~\cite{Smirnov2015} for more details). Even on the qualitative level one can notice the increase of the distance between the $\nu^+$ and $\nu^-$ components of the doublet with the resonance frequency, and the accompanying 'fading out' of the $\nu^+$ mode. Both effects are in agreement with the interacting spinon picture that can be inferred from Fig.~\ref{FIG:lines}(c) and Eq.~\eqref{EQ:complexDoublet}. For $u=0$, the intensities of two modes are equal and the splitting is field independent, determined by the magnitude of the DM vector only, as previously found in the low-field range below $1$~T~\cite{Smirnov2015}.

These observations can be further quantified by fitting measured ESR spectra with three overlapping Lorentzian functions to extract the precise intensities and resonant fields~\cite{SM}. Results of this new data analysis presented in Fig.~\ref{FIG:constantfits} show a striking quantitative agreement with the interacting spinon theory. The deviation between the data and the noninteracting spinon approximation \eqref{EQ:simpleDoublet} is clearly visible in the frequency-field
diagrams of Figs.~\ref{FIG:constantfits}(a),(b). While being completely unexplainable within the previous noninteracting approximation \eqref{EQ:simpleDoublet}, all the 'deviant' effects -- the upward
deflection and the fading of the upper $\nu^+$ mode, and the restoration of the Larmor precession for the lower $\nu^-$ mode -- are readily
explained by the new interacting spinon expression~(\ref{EQ:complexDoublet}),(\ref{EQ:deltadef}). The upward shift of the $\nu^+$ mode is the consequence of the gap
$\Delta$ \eqref{EQ:deltadef} growing with the field. The same effect is responsible for the upward approach of the $\nu^{-}$ mode towards the
Larmor frequency, as is seen in Fig.~\ref{FIG:constantfits}.

Fitting the $\nu(H)$ data to Eq.\eqref{EQ:complexDoublet}, we find an excellent agreement between the experiment and the interacting-spinon theory
for the value $\delta = 0.12\pm0.005$. By Eq.~\eqref{EQ:deltadef0}, this means that the backscattering interaction constant is $\hbar u/a=48.8\pm2$~K. This is a strong interaction
indeed, it corresponds to $2.38\pm0.1$ in the exchange coupling $J$ units. What matters, however, is that $u$ enters Eqs.~\eqref{EQ:complexDoublet} and \eqref{EQ:deltadef} only via $\delta$, which is quite small. This smallness implies the consistency of the made
theoretical assumptions. This is further confirmed by the spinon mean-free path estimate~\cite{LB-Egger2001}, and analysis of $u$ in terms of the RG approach~\cite{Lukyanov1998}~\cite{SM}.
Note that DM interaction strength $D$, while being an independent parameter of the fit, is actually unchanged compared to
the previous estimate $0.26\pm0.01$~K~\cite{Smirnov2015}. Its value is fixed by the zero field splitting $\pi D/2$ which at zero magnetization is
not affected by the interaction $u$.

The {\em intensity ratio} $I^{+}_\nu(H^{+})/I^{-}_\nu(H^{-})$ at a fixed frequency $\nu$ (with $H^{\pm}$ being the resonant fields of the corresponding modes) is another quantity that can be determined both theoretically and experimentally.
The theory~\cite{Keselman2020} predicts:

\begin{equation}
\label{EQ:intensity2}
    \frac{I^{+}_\nu(H^{+})}{I^{-}_\nu(H^{-})}=\frac{\sqrt{(h\nu\delta)^2+((1-\delta^2)\pi D/2)^2}-h\nu\delta}{\sqrt{(h\nu\delta)^2+((1-\delta^2)\pi D/2)^2}+h\nu\delta}.
\end{equation}

This parameter-free comparison is shown in Fig.~\ref{FIG:constantfits}(c). We find an excellent agreement between all our datasets and the theory
\eqref{EQ:intensity2} for the derived value $\delta=0.12$.
Notice that without the backscattering interaction $u$, i.e. for $\delta=0$, the ratio would be just $1$ for all frequencies.\footnote{Note again that \eqref{EQ:intensity2} describes the ratio of intensities of modes $\nu^\pm$ measured at the same fixed frequency but for different resonant fields $H^\pm$ while Figures \ref{FIG:lines}(b,c) illustrate relative intensities of $\nu^\pm$ modes for the fixed magnetic field. See~\cite{SM}~Sec.~IV.C.}

Thus, the relative
attenuation of the $\nu^{+}$ mode represents an additional confirmation of the validity of the interacting spinon description of \KCSOB.

To summarize, the observed field evolution of the ESR spectrum is very well explained by the model
of interacting spinons. The normally well-hidden backscattering interaction turns out to be a crucial ingredient for both {\em qualitative and  quantitative} description of the data.
The obtained value of the spinon backscattering interaction $u \simeq1.5 v_{F}\simeq 3.5 \cdot 10^5$ cm/s is of the order of spinon velocity
$v_F$. Experimental confirmation of the importance of interactions between spinons reveals a genuine
Fermi-liquid-like (in contrast to a Fermi-gas-like) behavior of the quasiparticles constituting the spin chain ground state. Dynamic small09-SM-Kirill-os-momentum response
of the quantum spin chain demonstrates an amazing similarity with an electron liquid and, in particular, Silin spin waves in nonferromagnetic
conductors~\cite{Silin1958,Silin1959}. This result paves way to spectroscopic investigations of more complex quantum spin liquids, including higher-dimensional
ones~\cite{LuoLakeMei_PRL_2018_SpinonQSLESR,BalentsStarykh_PRB_2020_SpinonsU1ESR}.

\acknowledgments

The work at ETH Z\"{u}rich has been supported by the SNSF Division II. We would like to thank Dr. Manuel H\"{a}lg and Dr. Wolfram Lorenz for their
involvement at the early stage of the project. The work of R.B.W. and O. A. S. is supported by the National Science Foundation CMMT program under
Grant No. DMR-1928919. The work at  Kapitza Institute (experiments, data processing and data analysis) has been supported by the Russian Science
Foundation Grants No. 17-12-01505 and 22-12-00259. O.A.S. would like to thank Anna Keselman and Leon Balents for the collaboration on the interacting spinons
project which helped him to get through the covid lockdown and provided theoretical foundation for the current investigation.

\bibliography{refs-spinons_v301}

\ifarXiv
    

\section*{Supplementary material}

# Supplemental Material for "Electron Spin Resonance of the Interacting Spinon Liquid"

Kirill Yu. Povarov,[1, *] Timofei A. Soldatov,[2] Ren-Bo Wang,[3]
Andrey Zheludev,[1] Alexander I. Smirnov,[2, †] and Oleg A. Starykh[3, ‡]

[1]*Laboratory for Solid State Physics, ETH Zürich, 8093 Zürich, Switzerland*
[2]*P. L. Kapitza Institute for Physical Problems RAS, 119334 Moscow, Russia*
[3]*Department of Physics and Astronomy, University of Utah, Salt Lake City, Utah 84112, USA*
(Dated: April 4, 2022)

In the Supplemental Material we discuss in detail the theoretical concepts related to the spin chains with uniform Dzyaloshinskii–Moriya interaction, their description in terms of interacting fermions, predictions for the ESR observables, Renormalization Group predictions, and estimates of the spinon mean-free path in the presence of backscattering interaction. We also give additional details of the used experimental and data analysis procedures, and compare the results described in the main text with the additional theoretical predictions.

## CONTENTS

---

* povarovk@phys.ethz.ch
† smirnov@kapitza.ras.ru
‡ starykh@physics.utah.edu

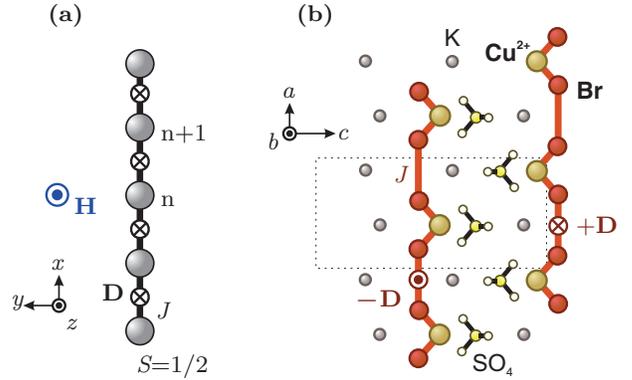

SUPP. FIG. 1. $K_2CuSO_4Br_2$ structure and the geometry notation used in the paper. (a) The cartoon representation of the spin chain described by the Hamiltonian (S.1). (b) $ac$ plane projection of $K_2CuSO_4Br_2$ crystal structure with the Cu-Br-Br-Cu chains running along $a$ and the DM vectors pointing along $b$ (the two-fold rotation axis of the structure). The key correspondances of the theoretical model to the crystollagraphic coordinate system of the actual material are: $x \to a$ and $z \to b$.

## I. THE MODEL AND THE UNITARY ROTATION

Consider a one-dimensional (1D) antiferromagnetic spin-$\frac{1}{2}$ Heisenberg chain with a uniform Dzyaloshinskii-Moriya (DM) interaction $\mathbf{D} = (0,0,-D)$ in the presence of an external magnetic field $\mathbf{H} = (0,\ 0,\ H)$ — the geometry reflected in Supp. Fig. 1 [1–4]

$$\mathcal{H} = J\sum_j (\hat{S}_j^x \hat{S}_{j+1}^x + \hat{S}_j^y \hat{S}_{j+1}^y + \hat{S}_j^z \hat{S}_{j+1}^z)$$
$$- D\sum_j (\hat{S}_j^x \hat{S}_{j+1}^y - \hat{S}_j^y \hat{S}_{j+1}^x) - \sum_j g\mu_B H \hat{S}_j^z. \quad \text{(S.1)}$$

For the considered parallel arrangement, $\mathbf{H} \parallel \mathbf{D}$, an important general consideration is possible already on the level of the lattice Hamiltonian. Namely, we rotate

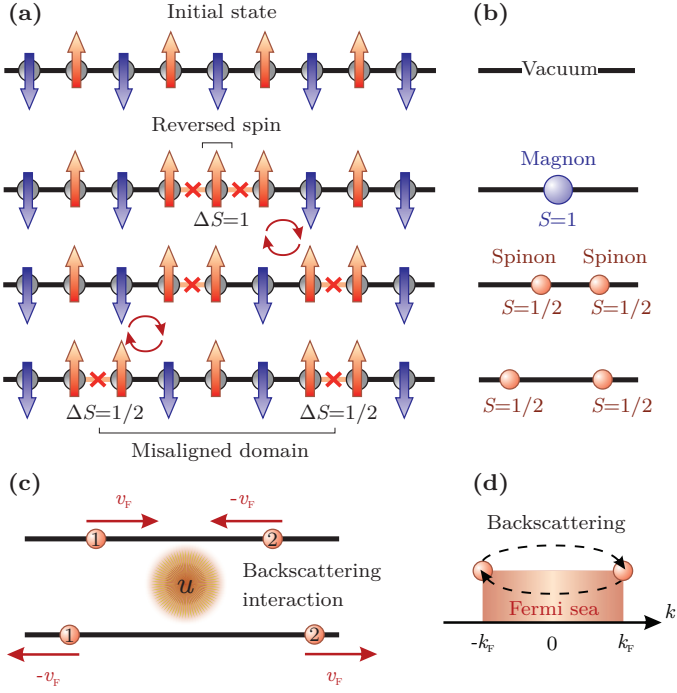

SUPP. FIG. 2. Fractionalization of excitations in one-dimensional antiferromagnets and their bacsckattering interaction. (a) Cartoon shows how a single flipped spin in a 1D antiferromagnetic chain 'breaks' into two domain walls at no energy cost. The total spin of the state remains 1 and the amount of broken bonds (red crosses) is always two. (b) The same process in the quasiparticle language. The antiferromagnetic state corresponds to the vacuum with no particles present, and a flipped spin corresponds to a single $S=1$ magnon excitation. Subsequently this particle decays into a pair of $S=1/2$ spinon excitations. (c) A real-space cartoon of bacsckattering interaction between the spinons. (d) Momentum space representation of backscattering process in a 1D Fermi system.

spins about $z$-axis as

$$\hat{S}_n^+ = \hat{\tilde{S}}_n^+ e^{-iq_{\rm DM} na}, \hat{S}_n^z = \hat{\tilde{S}}_n^z \quad (S.2)$$

where

$$q_{\rm DM} = \tan^{-1}(D/J)/a \approx \frac{D}{Ja} \quad (S.3)$$

and $x = na$ is the coordinate of the $n$-th spin along the chain, in units of the lattice spacing $a$.

The Hamiltonian (S.1) transforms into the XXZ form

$$\tilde{\mathcal{H}} = \sqrt{J^2+D^2} \sum_n \Big[\frac{1}{2}(\hat{\tilde{S}}_n^+ \hat{\tilde{S}}_{n+1}^- + \hat{\tilde{S}}_n^- \hat{\tilde{S}}_{n+1}^+)$$
$$+ \frac{J}{\sqrt{J^2+D^2}} \hat{\tilde{S}}_n^z \hat{\tilde{S}}_{n+1}^z \Big] - g\mu_B H \sum_n \hat{\tilde{S}}_n^z. \quad (S.4)$$

Observe that (S.4) describes a chain *without* the DM interaction with the slightly modified exchange interaction $\tilde{J} = \sqrt{J^2+D^2} \approx J + D^2/(2J)$ and the small easy-plane anisotropy $\tilde{\Delta} = \frac{J}{\sqrt{J^2+D^2}} \approx 1 - D^2/(2J^2)$. To the leading order in $D/J \ll 1$, these quadratic in $D/J$ deviations can be neglected.

The most important consequence of the simple transformation (S.2) is that the dynamic structure factor $\mathcal{S}^{+-}(q,\omega)$ of the Hamiltonian (S.1),

$$\mathcal{S}^{+-}(q,\omega) = \sum_n \int dt e^{i\omega t} e^{-iqna} \langle \hat{S}_n^+(t) \hat{S}_0^-(0)\rangle_{\mathcal{H}}, \quad (S.5)$$

where the expectation value $\langle ...\rangle_{\mathcal{H}}$ is taken with the respect to the equilibrium density matrix of the Hamiltonian $\mathcal{H}$, reduces to that of the rotated Hamiltonian (S.4),

$$\tilde{\mathcal{S}}^{+-}(q+q_{\rm DM},\omega) = \sum_n \int dt e^{i\omega t} e^{-iqna}$$
$$\times \langle e^{-iq_{\rm DM} na} \hat{\tilde{S}}_n^+(t) \hat{\tilde{S}}_0^-(0)\rangle_{\tilde{\mathcal{H}}}, \quad (S.6)$$

but with the *boosted* momentum $q+q_{\rm DM}$.

In particular, the ESR response of the original model (S.1), which is proportional to $\mathcal{S}^{+-}(0,\omega)$, is seen to be given by that of the essentially ideal spin chain (up to the omitted $D^2/J^2$ terms) $\tilde{\mathcal{S}}^{+-}(q_{\rm DM},\omega)$, evaluated at the *finite momentum* $q_{\rm DM} = D/(Ja)$.

This crucial feature of the spin chain with the uniform DM interaction turns the standard ESR measurement into the finite momentum probe and allows us to explore details of the small-momentum response of the spin-1/2 chain in the magnetic field with the accuracy greatly exceeding that of the inelastic neutron scattering experiments.

The above-mentioned relations also apply to the transverse dynamical susceptibility, defined by the retarded Green's function of the spin operators $S_n^+$ and $S_0^-$,

$$\chi^\pm(q,\omega) = -i \sum_n \int_0^\infty dt e^{i\omega t} e^{-iqna} \langle [S_n^+(t), S_0^-(0)]\rangle_{\mathcal{H}}. \quad (S.7)$$

It is connected with the dynamic structure factor by the familiar Fluctuation Dissipation theorem,

$$\mathcal{S}^{+-}(q,\omega) = -2(n(\omega)+1){\rm Im}[\chi^{+-}(q,\omega)]. \quad (S.8)$$

Here $n(\omega) = 1/(e^{\omega/T}-1)$ is the Bose function so that in the zero-temperature limit, $T\to 0$, the right-hand-side of (S.8) is non-zero only for $\omega > 0$.

The equivalence of the structure factors (S.5) and (S.6) translates into that of the susceptibilities,

$$\chi^\pm(q,\omega) = \tilde{\chi}^\pm(q+q_{\rm DM},\omega), \quad (S.9)$$

where $\tilde{\chi}^\pm(q,\omega)$ is the transverse susceptibility of the chain described by $\tilde{\mathcal{H}}$ (S.4) (equivalently, by Eq. (1) with no DM term, $D=0$).

It is also easy to see that for the transverse susceptibility for the opposite, $\mp$, circulation,

$$\chi^{\mp}(q,\omega) = -i \sum_n \int_0^\infty dt e^{i\omega t} e^{-iqna} \langle [S_n^-(t), S_0^+(0)] \rangle_{\mathcal{H}}, \tag{S.10}$$

the DM-induced shift occurs in the *opposite* direction,

$$\chi^{\mp}(q,\omega) = \tilde{\chi}^{\mp}(q - q_{\text{DM}}, \omega). \tag{S.11}$$

Finally, the longitudinal susceptibility does not experience the DM-induced shift of $q$ at all, $\chi^{zz}(q,\omega) = \tilde{\chi}^{zz}(q,\omega)$.

## II. FERMION MEAN-FIELD THEORY OF THE SPIN-1/2 HEISENBERG CHAIN

### A. Fermionic mapping

Here we sketch simple mean-field theory of the spin chain following [5] and [6, 7]. This serves to develop intuition for the more sophisticated interacting spinon description which is employed in the paper.

Start by writing spin operator on site $n$ in terms of Abrikosov fermion $c_{n\beta}$

$$\hat{S}_n^a = \frac{1}{2} \hat{c}_{n\alpha}^\dagger \sigma_{\alpha\beta}^a \hat{c}_{n\beta}, \tag{S.12}$$

where $\alpha, \beta \in (\uparrow, \downarrow)$ are spin indices and $a = x, y, z$ denotes spin component. The convention is that repeated spin indices $\alpha, \beta...$ are summed over, and therefore we do not write explicit sums over them. The fermions satisfy standard anticommutation relation $\{\hat{c}_{n\alpha}, \hat{c}_{m\beta}^\dagger\} = \delta_{n,m}\delta_{\alpha,\beta}$. This representation is exact as long as there is exactly one fermion on every site (the constraint condition)

$$\sum_{\alpha=\uparrow,\downarrow} \hat{c}_{n\alpha}^\dagger \hat{c}_{n\alpha} = 1. \tag{S.13}$$

This representation converts Heisenberg Hamiltonian into that of auxilliary fermions with quartic interaction

$$\hat{\mathcal{H}} = J \sum_n \sum_{a=x,y,z} \hat{S}_n^a \hat{S}_{n+1}^a = \frac{J}{4} \sum_n \sum_{a=x,y,z} \hat{c}_{n\alpha}^\dagger \sigma^a \hat{c}_{n\beta} \, \hat{c}_{n+1\gamma}^\dagger \sigma^a \hat{c}_{n\delta}, \tag{S.14}$$

but Fierz identity

$$\sum_a \sigma_{\alpha\beta}^a \sigma_{\gamma\delta}^a = 2\delta_{\alpha\delta}\delta_{\beta\gamma} - \delta_{\alpha\beta}\delta_{\gamma\delta} \tag{S.15}$$

helps to simplify it. Namely,

$$\hat{\mathcal{H}} = \frac{J}{4} \sum_n \left( 2\hat{c}_{n\alpha}^\dagger \hat{c}_{n\beta} \, \hat{c}_{n+1\beta}^\dagger \hat{c}_{n+1\alpha} - \hat{c}_{n\alpha}^\dagger \hat{c}_{n\alpha} \hat{c}_{n\gamma}^\dagger \hat{c}_{n\gamma} \right). \tag{S.16}$$

Notice that the last term in the above equation is, according to the constraint (S.13), the product of identity operators and contributes trivial constant to $H$, which we omit in the following. Therefore the Hamiltonian

$$\hat{\mathcal{H}} = -\frac{J}{2} \sum_n \hat{c}_{n\alpha}^\dagger \hat{c}_{n+1\alpha} \hat{c}_{n+1\beta}^\dagger \hat{c}_{n\beta} \tag{S.17}$$

is now written in terms of the bond operator $\hat{\mathcal{B}}_{n,n+1} = \hat{c}_{n\alpha}^\dagger \hat{c}_{n+1\alpha} = \hat{c}_{n\uparrow}^\dagger \hat{c}_{n+1\uparrow} + \hat{c}_{n\downarrow}^\dagger \hat{c}_{n+1\downarrow}$ and its conjugate $\hat{\mathcal{B}}_{n,n+1}^\dagger = \hat{\mathcal{B}}_{n+1,n}$.

A standard "cartoon" for the fractionalized fermionic excitations in the Ising chain case can be found in Supp. Fig. 2. Here a spinon can be visualized as a kink (domain wall) separating two opposite Néel domains. Therefore such an excitation can not be represented by any finite number of local spin operators.

### B. Mean Field description

Up to now the manipulations are exact. Now we proceed with the mean field approximation by introducing position-independent and real vacuum expectation value [5]

$$\mathcal{D} = \langle \hat{c}_{n\uparrow}^\dagger \hat{c}_{n+1\uparrow} + \hat{c}_{n\downarrow}^\dagger \hat{c}_{n+1\downarrow} \rangle \neq 0 \tag{S.18}$$

and assuming that fluctuations of $\hat{\mathcal{B}}_{n,n+1}$ about $\mathcal{D}$ are *small*. This assumption, consistency of which is checked aposteriori [5], allows us to neglect quadratic in this smallness terms $(\hat{\mathcal{B}}_{n,n+1} - \mathcal{D})(\hat{\mathcal{B}}_{n+1,n} - \mathcal{D}) \to 0$ and obtain

$$\hat{\mathcal{H}} = -\frac{J}{2}\sum_n(\mathcal{D}+\hat{\mathcal{B}}_{n,n+1}-\mathcal{D})(\mathcal{D}+\hat{\mathcal{B}}_{n+1,n}-\mathcal{D}) \approx -\frac{J}{2}\sum_n \mathcal{D}^2 + \mathcal{D}\Big(\hat{\mathcal{B}}_{n,n+1}-\mathcal{D}+\hat{\mathcal{B}}_{n+1,n}-\mathcal{D}\Big)$$

$$= \frac{J}{2}\sum_n \mathcal{D}^2 - \frac{J}{2}\sum_n \mathcal{D}\Big(\hat{\mathcal{B}}_{n,n+1}+\hat{\mathcal{B}}_{n+1,n}\Big) \to -\frac{J\mathcal{D}}{2}\sum_n \big(\hat{c}^\dagger_{n\alpha}\hat{c}_{n+1\alpha}+\hat{c}^\dagger_{n+1\alpha}\hat{c}_{n\alpha}\big), \qquad (S.19)$$

where in the last step we dropped unessential constant $\propto \mathcal{D}^2$ from the previous one.

The last expression in (S.19) describes non-interacting fermions. It is now straightforward to Fourier transform to the $k$-space

$$\hat{c}_{n\alpha} = \frac{1}{\sqrt{N}}\sum_k e^{-ikn}\hat{c}_{k\alpha} \qquad (S.20)$$

and also to add magnetic field via the Zeeman interaction term

$$-\sum_n g\mu_B H \hat{S}^z_n = -\frac{1}{2}g\mu_B H\sum_n(\hat{c}^\dagger_{n\uparrow}\hat{c}_{n\uparrow}-\hat{c}^\dagger_{n\downarrow}\hat{c}_{n\downarrow}) = -\frac{1}{2}g\mu_B H\sum_k(\hat{c}^\dagger_{k\uparrow}\hat{c}_{k\uparrow}-\hat{c}^\dagger_{k\downarrow}\hat{c}_{k\downarrow}). \qquad (S.21)$$

In this way we arrive at the Mean Field Hamiltonian of the spin chain in the magnetic field (here $\uparrow$ corresponds to $\alpha=+1$ and $\downarrow$ to $\alpha=-1$)

$$\hat{\mathcal{H}}_{\rm MF} = \sum_k \sum_{\alpha=\pm 1}\Big(-J\mathcal{D}\cos(k) - \frac{1}{2}g\mu_B H\alpha\Big)\hat{c}^\dagger_{k\alpha}\hat{c}_{k\alpha} \qquad (S.22)$$

### C. The Larmor mode and low-$q$ spinon continuum in applied field

Without the magnetic field the fermion band is half-filled (due to the constraint (S.13)) which means that its Fermi wavevector is $k_F = \pi/(2a)$. Finite magnetic field splits this degeneracy, resulting in different Fermi wavevectors for $\uparrow$- and $\downarrow$-spin fermions,

$$\langle \hat{S}^z\rangle = \frac{1}{2}(n_\uparrow - n_\downarrow) = \qquad (S.23)$$
$$\frac{1}{2}\int_{-\pi}^{\pi}\frac{dka}{2\pi}(n_\uparrow(k)-n_\downarrow(k)) = \frac{a}{2\pi}(k_{F\uparrow}-k_{F\downarrow})$$

where $\langle \hat{S}^z\rangle$ is the average spin angular momentum per site of the chain. Note that this implies that the spin angular momentum density $M$ is given by $\langle \hat{S}^z\rangle/a$. The number of $\uparrow$-fermions grows, $k_{F\uparrow}\to k_{F\uparrow}+\Delta k_F$, while that of $\downarrow$-fermions decreases, $k_{F\downarrow}\to k_{F\downarrow}-\Delta k_F$. Therefore

$$\Delta k_F = \pi M \qquad (S.24)$$

changes from 0 for the non-magnetized chain to the maximum possible value of $\pi/(2a)$ for the fully magnetized one ($\langle \hat{S}^z\rangle = 1/2$), when the $\downarrow$-fermion band is completely depopulated.

The above picture provides a very intuitive way of understanding the structure of low-$q$ continuum depicted in Fig. 2a of the main text. Naturally, the susceptibilities $\chi^\pm$ and $\chi^\mp$ correspond to spin-flip transitions $\downarrow,\uparrow$-to-$\uparrow,\downarrow$ in the neighborhood of the zero-field Fermi momentum $\pi/2a$, and vice versa. A very special transition, with the zero momentum change $q=0$ and the energy transfer of $g\mu_B H$, is the Larmor resonance. Since $g\mu_B H$ is exactly the Zeeman energy splitting of the Fermi surface, many fermions ($4\Delta k_F$ of them) contribute to it, as Supp. Fig. 3(left) shows. This is true for both linear and non-linear spinon dispersions, e.g. such as a cosine one in (S.22). Generally, this is not the case for $q\ne 0$ transitions and the non-linear dispersion of spinons, Supp. Fig. 3(left).

Details of the non-interacting spinon continuum in Figs. 3 and 4 are understood from the transverse susceptibility

$$\chi^\pm(x,t) = -i\Theta(t)\langle[S^+(x,t),S^-(0,0)]\rangle = -i\Theta(t)\frac{1}{N^2}\sum_{k,k'}\sum_{k_1,k_2}e^{i\epsilon_k-\epsilon_{k'}-g\mu_B H)t}e^{-i(k-k')x}\langle[\hat{c}^\dagger_{k\uparrow}\hat{c}_{k'\downarrow},\hat{c}^\dagger_{k_1\downarrow}\hat{c}_{k_2\uparrow}]\rangle, \qquad (S.25)$$

where we parameterized spin-$\alpha$ spinon dispersion as $\epsilon_k - \frac{1}{2}g\mu_B H\alpha$. Eq.(S.22) corresponds to $\epsilon_k = -J\cos(k)$ (we absorb $\mathcal{D}$ into $J$ for simplicity). Other dispersion relations are possible as well, and results for the linear and parabolic ones are presented in Supp. Fig. 4. The commutator in (S.25) is easily evaluated to be

$$\langle[\hat{c}_{k\uparrow}^\dagger \hat{c}_{k'\downarrow}, \hat{c}_{k_1\downarrow}^\dagger \hat{c}_{k_2\uparrow}]\rangle = \delta_{k',k_1}\delta_{k,k_2}\Big(f(\epsilon_k - g\mu_B H/2) - f(\epsilon_{k'} + g\mu_B H/2)\Big), \tag{S.26}$$

where $f$ is the Fermi function $f(x) = 1/(e^{x/T}+1) \to \Theta(-x)$ in the zero-temperature limit $T \to 0$. Fourier transforming

$$\chi^\pm(q,\omega) = \int dt \int dx e^{-iqx+i\omega t}\chi^\pm(x,t) = \frac{1}{N}\sum_k \frac{f(\epsilon_k - g\mu_B H/2) - f(\epsilon_{k'} + g\mu_B H/2)}{\omega - g\mu_B H + \epsilon_k - \epsilon_{k+q} + i0} \tag{S.27}$$

so that the imaginary part is

$$\mathrm{Im}\Big(\chi^\pm(q,\omega)\Big) = -\pi \int \frac{dk}{2\pi}\Big(\Theta(-\epsilon_k + g\mu_B H/2) - \Theta(-\epsilon_{k+q} - g\mu_B H/2)\Big)\delta(\omega - g\mu_B H + \epsilon_k - \epsilon_{k+q}). \tag{S.28}$$

Step-functions select two integration intervals: $(\pi/2 - \phi - q, \pi/2 + \phi)$ and $(-\pi/2 - \phi, -\pi/2 + \phi - q)$ where $\sin\phi = g\mu_B H/(2J)$. The first of these corresponds to the right-moving fermions while the second - to the left ones. Focusing on the right movers, set $k = \pi/2 + p - q/2$ so that the argument of the delta-function becomes $\omega - g\mu_B H - 2J\sin(q/2)\cos(p)$. The $p$-integration is simple and leads to the right-movers contribution to $\chi$

$$\mathrm{Im}\Big(\chi_R^\pm(q,\omega)\Big) = \frac{-1}{\sqrt{(2J\sin q/2)^2 - (\omega - g\mu_B H)^2}} \text{ where } 2J\sin(q/2)\cos(\phi+q/2) \leq \omega - g\mu_B H \leq 2J\sin(q/2) \tag{S.29}$$

This is seen as the upward branch in Supp. Fig.3(right) emerging from $g\mu_B H$. Contribution of the left-moving fermions has identical functional form but the frequency is bounded by $2J\sin(q/2)\cos(\phi - q/2) \leq \omega - g\mu_B H \leq 2J\sin(q/2)$. This is seen to be extremely narrow, essentially a single line, downward branch in Supp. Fig.3(right). $\mathrm{Im}\Big(\chi^\mp(q,\omega)\Big)$ is obtained from (S.29) by $H \to -H$.

Figure 4 shows examples of the spinon continua for different $\epsilon_k$. In particular, for the linear dispersion (which is the case of our Hamiltonian (2) in the main text) susceptibility (S.28) reduces to the sum of $\delta(\omega - g\mu_B H \pm v_F k)$ as described in the main text, see left panel of Supp. Fig. 4. Interactions (different from the backscattering one we consider in this work) do spread spinon continuum outside the narrow intervals obtained in (S.29) and below [3]. But most of the spectral weight remains to be sharply focused on the boundaries of the non-interacting spinon continuum, in a delta-function-like fashion.

Backscattering interaction $u$ changes this picture qualitatively, by shifting the spinon continuum up in energy and producing a spin-1 oscillatory collective mode of spinons [8], that originates from the Larmor frequency at $q = 0$. Its effect on the spin chain with the uniform DM interaction is described in the main text, while key technical steps are summarized in the App. III below.

## III. LOW-ENERGY MODEL

### A. Simplified notation

In the following part we are going to use the simplified notation. The following correspondences are used:

- Zeeman energy as the magnetic field equivalent $g\mu_B H \to B$
- Excitation energy $h\nu \to \omega$
- Susceptibility per unit of chain length $\chi_0 = \frac{(g\mu_B)^2}{2\pi v} = \frac{(g\mu_B)^2}{a\pi^2 J}$ is expressed as $1/\pi^2 J$ (corresponding to *single **spin*** susceptibility with respect to field $B$ defined above)

Effectively, in this simplified notation the lattice constant $a$, $g\mu_B$, and $\hbar$ are all set to 1.

### B. Model formulation

Neglecting small $\propto D^2$ anisotropy in (S.4) turns the Hamiltonian $\tilde{\mathcal{H}}$ into that of the ideal Heisenberg spin chain. A simple analytical expression for the dynamic spin susceptibility of this system was recently derived, and checked with extensive DMRG simulations, in Ref. 8. Here we summarize theoretical formulation of the problem that leads to the desired results for the transverse susceptibility $\tilde{\chi}^\pm$ and $\tilde{\chi}^\mp$.

The spin operators are approximated as [1, 4, 9]

$$\hat{\tilde{\mathbf{S}}}_n \to [\hat{\mathbf{J}}_L(x) + \hat{\mathbf{J}}_R(x) + (-1)^x \hat{\mathbf{N}}(x)], \tag{S.30}$$

where $x = na \to n$ is the spin position, $\hat{\mathbf{J}}_L$ and $\hat{\mathbf{J}}_R$ are the uniform spin densities at the left and right Fermi points (they are called as left/right spin-currents in theoretical literature), and $\hat{\mathbf{N}}$ is the staggered component of the local spin density (also known as $2k_F$ component). The low-energy effective Hamiltonian can be written as (this is known as the Sugawara form),

6

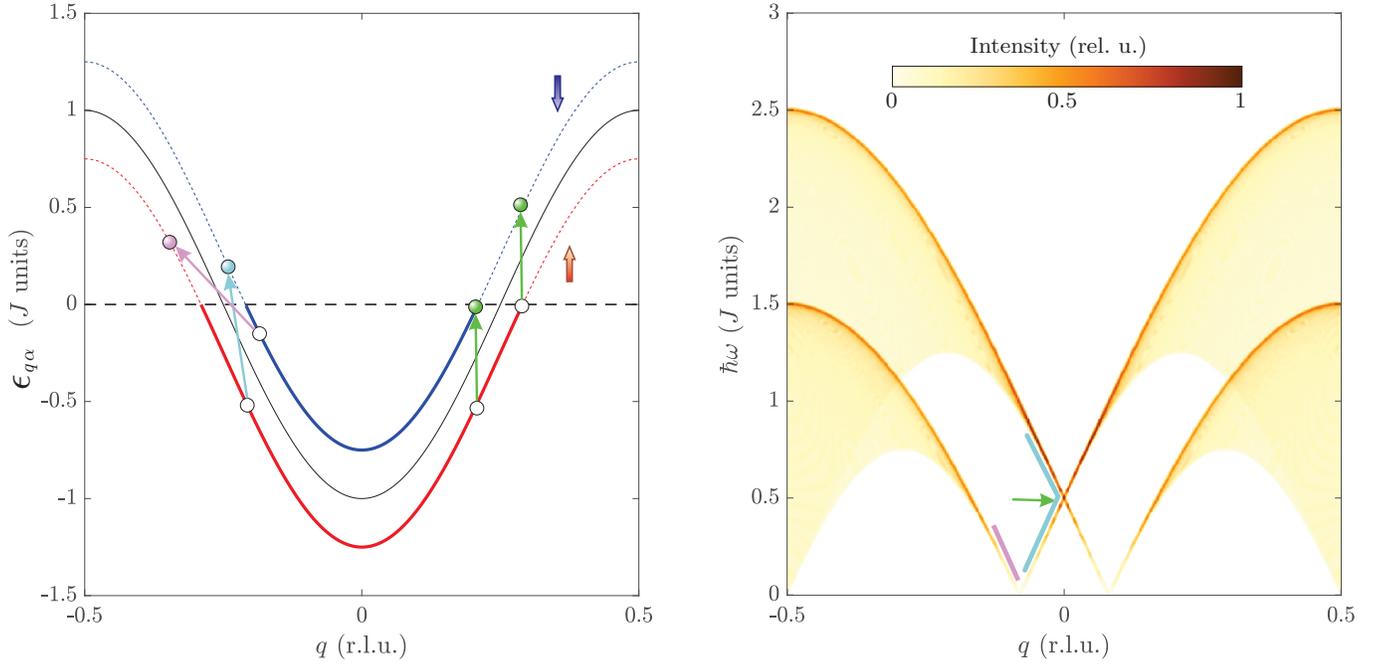

SUPP. FIG. 3. Left: spinon dispersions for up and down values of spin $\alpha$ in the applied field $g\mu_B H = 0.5J$. The typical spin-flip transitions scenarios with small momentum transfer $q$ are shown. Right: the resulting transverse two-spinon continuum. Sections of the spectra originating from the transitions illustrated by colored arrows in the left panel are highlighted by the same colors in the right panel.

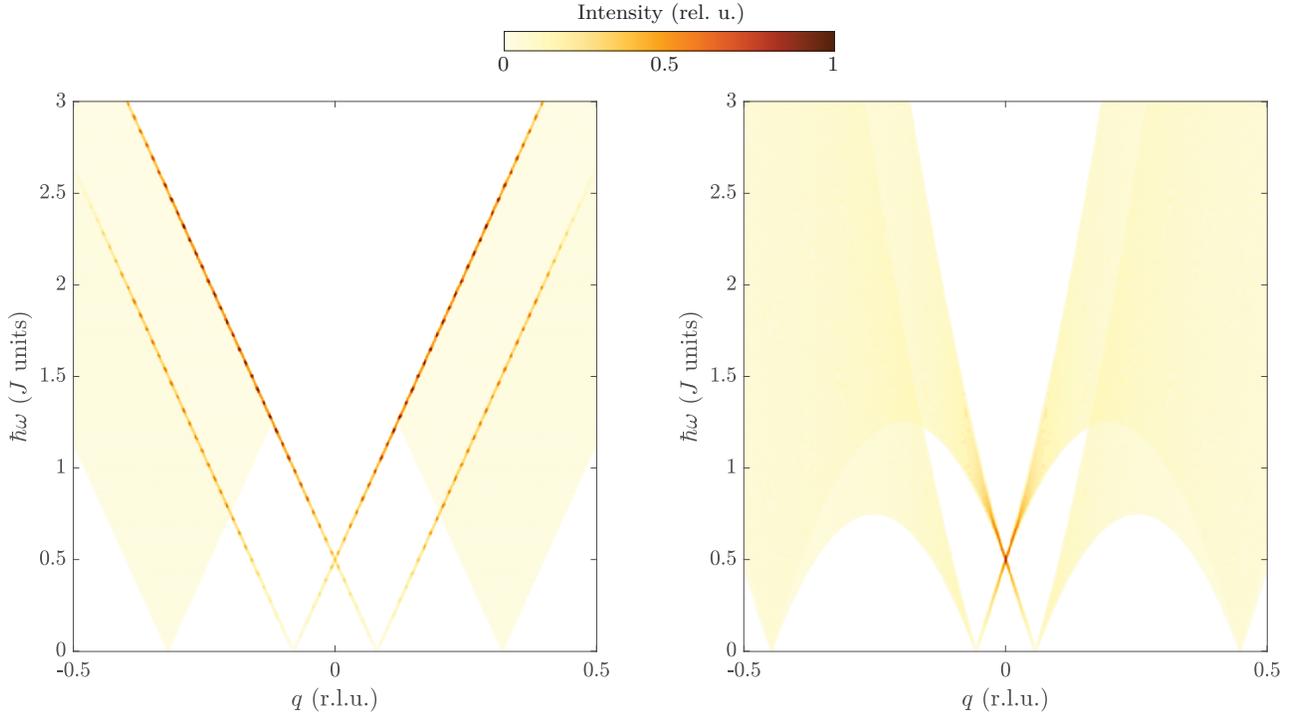

SUPP. FIG. 4. Two-fermion spin-flip continuum for the toy models with linear dispersion $\epsilon_q = J(q-1)$ (left) and parabolic dispersion $\epsilon_q = J(\frac{q^2}{2J} - 1)$ (right). Zeeman energy is $g\mu_B H = 0.5J$ in both cases. Observe that for small wavevectors the left panel of the figure is equivalent to Fig. 2b.

$$\hat{\mathcal{H}} = \hat{\mathcal{H}}_0 + \hat{V}_{\text{bs}} + \hat{\mathcal{H}}', \tag{S.31}$$

$$\hat{\mathcal{H}}_0 = \frac{2\pi}{3} v_F \int dx \, : \hat{\mathbf{J}}_R \cdot \hat{\mathbf{J}}_R + \hat{\mathbf{J}}_L \cdot \hat{\mathbf{J}}_L :, \tag{S.32}$$

$$\hat{V}_{\text{bs}} = -u \int dx \, : \hat{\mathbf{J}}_R \cdot \hat{\mathbf{J}}_L : \tag{S.33}$$

$$\hat{\mathcal{H}}' = -B \int dx \, (\hat{J}_R^z + \hat{J}_L^z), \tag{S.34}$$

where $v_F = \pi J/2$ is the speed of the low-energy excitations, columns : : denote normal ordering (which places creation operators to the left of annihilation ones).

The amplitude of the backscattering interaction, cartoon of which can be found in Supp. Fig. 2c,d, is denoted by $u$. It plays key role in our study. It is known as a *running* coupling constant within the Renormalization Group (RG) technique [9–11]. Due to the dependence on spinon distribution functions, this marginally irrelevant (in the RG sense) interaction $u(T, H)$ is weak (typically, logarithmic) function of the temperature $T$ and magnetic field $H$, that vanishes in the limit $T, H \to 0$. See equations (S.59) and (S.60) illustrating this behavior. However, for any finite $T$ and/or $H$ it is finite and constitutes the main inter-spinon interaction in the problem.

Note that the magnetization operator $\hat{\mathbf{M}} = \hat{\mathbf{J}}_R + \hat{\mathbf{J}}_L$ is the sum of the right and left spin currents. Correlation function of $\hat{\mathbf{M}}$ determines the uniform susceptibility. Its expectation value is the magnetization itself, $M = \langle \hat{J}_R^z + \hat{J}_L^z \rangle$. The staggered susceptibility, which is best probed by the neutron scattering, is determined by the correlation function of $\hat{\mathbf{N}}$.

Hamiltonian (S.31) can be re-written in terms of spin-1/2 Dirac fermions $\psi_{R,s}, \psi_{L,s}$ [9].

Then the spin current reads $\hat{\mathbf{J}}_{R/L} = \frac{1}{2}\hat{\psi}^\dagger_{R/L}\boldsymbol{\sigma}\hat{\psi}_{R/L}$. The notation is that $\hat{\psi}_R$ without spin subindex $s = \uparrow, \downarrow$ denotes the two-component spinor $\hat{\psi}_R = (\hat{\psi}_{R\uparrow}, \hat{\psi}_{R\downarrow})^{\text{T}}$. The full Hamiltonian is $\hat{\mathcal{H}} = \hat{\mathcal{H}}_0 + \hat{\mathcal{H}}' + \hat{V}_{\text{bs}}$ where the free part $\hat{\mathcal{H}}_0$ describes fermion kinetic energy and $\hat{\mathcal{H}}'$ collects DM and Zeeman interactions

$$\hat{\mathcal{H}}_0 + \hat{\mathcal{H}}' = \int dx \left[ v_F \left( \hat{\psi}^\dagger_R (-i\partial_x) \hat{\psi}_R + \hat{\psi}^\dagger_L (i\partial_x) \hat{\psi}_L \right) - \frac{B}{2} (\psi^\dagger_R \sigma^z \psi_R + \psi^\dagger_L \sigma^z \psi_L) \right] \tag{S.35}$$

The backscattering interaction is the sum of transverse and longitudinal parts, $\hat{V}_{\text{bs}} = \hat{V}_1 + \hat{V}_2$, where

$$\hat{V}_1 = -\frac{u}{2}\int dx \, : \left( \hat{J}_R^+ \hat{J}_L^- + \hat{J}_R^- \hat{J}_L^+ \right) := -\frac{u}{2}\int dx \, \left( \hat{\psi}^\dagger_{R\uparrow}\hat{\psi}_{R\downarrow}\hat{\psi}^\dagger_{L\downarrow}\hat{\psi}_{L\uparrow} + \hat{\psi}^\dagger_{R\downarrow}\hat{\psi}_{R\uparrow}\hat{\psi}^\dagger_{L\uparrow}\hat{\psi}_{L\downarrow} \right) : \tag{S.36}$$

and

$$\hat{V}_2 = -u \int dx \, : \hat{J}_R^z \hat{J}_L^z := -\frac{u}{4}\int dx \, : (\hat{\psi}^\dagger_{R\uparrow}\hat{\psi}_{R\uparrow} - \hat{\psi}^\dagger_{R\downarrow}\hat{\psi}_{R\downarrow})(\hat{\psi}^\dagger_{L\uparrow}\hat{\psi}_{L\uparrow} - \hat{\psi}^\dagger_{L\downarrow}\hat{\psi}_{L\downarrow}) : . \tag{S.37}$$

Backscattering interaction $\hat{V}_{\text{bs}}$ strongly alters the two-spinon continuum (left panel of Supp. Fig.4 and Fig. 2a) and is responsible for its transformation from Fig. 2b to Fig. 2c in the case of the interacting spinon liquid with $u \neq 0$.

Conversely, the fermion representation of the Heisenberg chain introduced here clearly shows that the theory without the backscattering interaction, $u = 0$, actually describes free non-interacting Dirac fermions, as given by (S.35).

### C. Connection with the abelian bosonization

For completeness, here we write down abelian form of equations (S.35), (S.36) and (S.37). Chapter 18 of Ref. [9], see also Appendix A in [4], shows that

$$\hat{\mathcal{H}}_0 + \hat{V}_2 = \int dx \frac{1}{2}[v_F K_s (\partial_x \theta_s)^2 + \frac{v_F}{K_s}(\partial_x \varphi_s)^2], \tag{S.38}$$

$$K_s = 1 + \frac{u}{4\pi v_F},$$

$$\hat{\mathcal{H}}' = -\int dx \frac{B}{\sqrt{2\pi}} \partial_x \varphi_s, \tag{S.39}$$

$$\hat{V}_1 = -\frac{u}{4\pi^2} \int dx \cos \sqrt{8\pi} \varphi_s. \tag{S.40}$$

Observe that longitudinal part of the backscattering interaction, term $\hat{V}_2$, adds to the non-interacting Hamiltonian $\hat{\mathcal{H}}_0$. This apparent simplicity, however, hides spin-rotational invariance of the theory. In the $u \to 0$ limit, which takes place at $T = 0$ and for $B = 0$, the Hamiltonian reduces to the non-interacting one, with $K_s = 1$ and $\hat{V}_1 = 0$. For the purpose of our study, the fermion

formulation of the theory in Section III B is much more convenient.

## IV. SUSCEPTIBILITY OF THE SPIN CHAIN FOR H ∥ D

### A. Transverse susceptibility

Ref. 8 finds
$$\tilde{\chi}^{\pm}(q,\omega) = \chi_0 \left( \frac{A_+(q)}{\omega - \omega_+(q) + i0} + \frac{A_-(q)}{\omega - \omega_-(q) + i0} \right),$$
$$A_{\pm}(q) = \frac{B}{1-\delta} \pm \frac{(1+\delta)v_F^2 q^2 - \delta(B/(1-\delta))^2}{\sqrt{\Delta^2 + \tilde{v}^2 q^2}},$$
$$\omega_{\pm}(q) = B + \Delta \pm \sqrt{\Delta^2 + \tilde{v}^2 q^2}. \quad \text{(S.41)}$$

Here we denote
$$\delta = \frac{1}{2} u \chi_0 = \frac{u}{4\pi v_F},$$
$$\tilde{v} = v_F \sqrt{1 - u^2 \chi_0^2/4} = v_F \sqrt{1-\delta^2},$$
$$\Delta = \frac{1}{2} u M = \frac{u \chi_0 B}{2(1-\delta)} = \frac{\delta B}{1-\delta}. \quad \text{(S.42)}$$

Therefore, the ESR experiment measures $\text{Im}[\tilde{\chi}^{\pm}(q_{\text{DM}},\omega)]$ and $\text{Im}[\tilde{\chi}^{\mp}(-q_{\text{DM}},\omega)]$, where $\tilde{\chi}^{\mp}$ is obtained from $\tilde{\chi}^{\pm}$ by changing the sign of the momentum $q$ and flipping the magnetic field $B \to -B$ so that magnetization changes sign too, $M \to -M$ (Onsager's relation):
$$\tilde{\chi}^{\mp}(q,\omega)|_B = \tilde{\chi}^{\pm}(-q,\omega)|_{-B}. \quad \text{(S.43)}$$

Note that throughout the manuscript we are consistently denoting the susceptibility of the ideal spin-1/2 chain by $\tilde{\chi}(q,\omega)$, the notation without the tilde, $\chi(q,\omega)$, is reserved for the susceptibility of the chain with DM interaction present, that is for the $K_2CuSO_4Br_2$ Hamiltonian (S.1). Of course, the two susceptibilities are related by the simple equation (S.9).

### B. ESR Modes

Now we specialize (S.42) to the ESR set-up studied in this paper. This is done by the simple substitution $\hbar v_F q_{\text{DM}} = \pi D/2$ as explained in the main text following Eq. (4). Thus
$$\omega_{\pm}(q_{\text{DM}}) = \frac{B}{1-\delta} \pm \sqrt{\left(\frac{\delta B}{1-\delta}\right)^2 + (1-\delta^2)\left(\frac{\pi}{2}D\right)^2} \quad \text{(S.44)}$$

The residues simplify similarly
$$A_{\pm}(q_{\text{DM}}) = \frac{B}{1-\delta} \pm \frac{(1+\delta)(\pi D/2)^2 - B^2 \delta/(1-\delta)^2}{\sqrt{\left(\frac{\delta B}{1-\delta}\right)^2 + (1-\delta^2)\left(\frac{\pi}{2}D\right)^2}} \quad \text{(S.45)}$$

The intensity of the signal follows from (S.82) and (S.41),
$$\text{Im}[\tilde{\chi}^{\pm}(q_{\text{DM}},\omega)] = -\pi \chi_0 \Big( A_+ \delta(\omega - \omega_+) + A_- \delta(\omega - \omega_-) \Big). \quad \text{(S.46)}$$

Observe that $\omega_+ > 0$ for all fields and therefore contributes to the ESR signal for all fields $B \in (0,\infty)$. The lower branch, $\omega_-$, becomes positive for $B > B_1$, see (S.50) below. Thus for that branch $B \in (B_1, \infty)$.

The total intensity of this polarization is therefore the sum of the two contributions
$$I^{+-}(\omega) = \frac{\pi}{2} H_{\text{rad}}^2 \left( \omega_+ A_+ \delta(\omega - \omega_+) + \omega_- A_- \delta(\omega - \omega_-) \right). \quad \text{(S.47)}$$

Correspondingly, the ratio of the intensities of the upper and lower ESR branches is given by
$$\frac{I^{+-}(\omega_+)}{I^{+-}(\omega_-)} = \frac{\omega_+(q_{\text{DM}})}{\omega_-(q_{\text{DM}})} \frac{A_+(q_{\text{DM}})}{A_-(q_{\text{DM}})}. \quad \text{(S.48)}$$

Next we turn to $\chi^{-+}$, the opposite circular polarization. Using (S.43)
$$\text{Im}[\tilde{\chi}^{\mp}(q_{\text{DM}},\omega)] = -\pi \chi_0 \Big( -A_+ \delta(\omega + \omega_+) - A_- \delta(\omega + \omega_-) \Big). \quad \text{(S.49)}$$

Since $\omega_+(q_{\text{DM}}) > 0$ for all $B, D$, the first term in the above does not contribute to the imaginary part for $\omega > 0$. The 2nd term does contribute, when the field is within the interval $B \in (0, B_1)$, so that $\omega(q_{\text{DM}}) \leq 0$. Here
$$B_1 = (1-\delta) \frac{\pi D}{2} \quad \text{(S.50)}$$

is the maximum field for the response in this polarization. It is also easy to check that $-A_-(q_{\text{DM}}) \geq 0$ for $B \in (0, B_1)$ and vanishes when $B_1$ is reached. The intensity of this branch is therefore
$$I^{-+}(\omega) = \frac{\pi}{2} H_{\text{rad}}^2 \omega_-(q_{\text{DM}}) \big( -A_-(q_{\text{DM}}) \big) \delta(\omega + \omega_-). \quad \text{(S.51)}$$

### C. ESR Intensity

Eq.(S.48) describes ratio of intensities at a given (fixed) magnetic field, when frequencies of the two modes are different. But the experiment is done at fixed resonator frequency by adjusting external magnetic field, i.e. the two branches contribute at the same frequency which occurs at different resonance fields for them. To account for this, we need to switch from sum of delta-functions $\delta(\omega - \omega_{\pm})$ to the one which depends on magnetic field $B$, $\delta(B - B_{\pm})$. Therefore, let
$$\omega = \frac{B_{\pm}}{1-\delta} \pm \sqrt{\left(\frac{\delta B_{\pm}}{1-\delta}\right)^2 + (1-\delta^2)(\pi D/2)^2}. \quad \text{(S.52)}$$

We find

$$B_\pm = \frac{\omega \mp S}{1+\delta}, \; S = \sqrt{\delta^2 \omega^2 + (1-\delta^2)^2 (\pi D/2)^2}. \quad (S.53)$$

By construction, $\omega_+(B_+) = \omega = \omega_-(B_-)$ and $B_- - B_+ = 2S/(1+\delta) > 0$. Next, we need to "solve" the delta function $\delta(\omega - \omega_\pm)$ by using $\omega_\pm(H) = \omega_\pm(B_\pm) + (B - B_\pm)\omega'_\pm$, where $\omega'_\pm \equiv (d\omega_\pm(B)/dB)|_{B=B_\pm}$. Then $\delta(\omega - \omega_\pm) = \delta(B - B_\pm)/|\omega'_\pm|$. Using this result and after considerable algebra we find that (S.47) is replaced by

$$I_\omega^{+-}(H) = \frac{1}{2} H_{\mathrm{rad}}^2 \omega \frac{\pi \chi_0}{1-\delta} \Big[ \frac{S+\delta\omega}{S} \delta(B-B_-) \\ + \frac{S-\delta\omega}{S}\delta(B-B_+)\Big]. \quad (S.54)$$

Therefore we get for the ratio of the intensities at the same frequency

$$\mathcal{R} = \frac{I_\omega^{+-}(B_+)}{I_\omega^{+-}(B_-)} = \frac{\sqrt{(\delta\omega)^2 + (1-\delta^2)^2 (\pi D/2)^2} - \delta\omega}{\sqrt{(\delta\omega)^2 + (1-\delta^2)^2 (\pi D/2)^2} + \delta\omega}. \quad (S.55)$$

This simple result has few nice features. For non-interacting spinons ($\delta = 0$) we obtain $\mathcal{R} = 1$, while for the interacting ones but *without* the DM interaction ($D = 0$) we have $\mathcal{R} = 0$ (since in this case one is probing at strictly $q = 0$). For both $\delta \neq 0$ and $D \neq 0$ we see that $\mathcal{R}$ is decreasing function of the frequency. All these limits make sense.

### D. Important formulas in full "experimental" notations

The frequencies of the $\pm$ ESR modes given by Eq. (S.44) are:

$$h\nu_\pm(H) = \frac{g\mu_B H}{1-\delta} \pm \sqrt{\Big(\frac{\delta}{1-\delta}g\mu_B H\Big)^2 + (1-\delta^2)\Big(\frac{\pi}{2}D\Big)^2}. \quad (S.56)$$

The corresponding spectral weights (S.45) are:

$$A_\pm(H) = \frac{1}{\pi^2 a J (1-\delta)}\Big[g\mu_B H + \\ \pm \frac{(\pi D/2)^2 - (g\mu_B H)^2 \delta/(1-\delta)}{\sqrt{\Big(\frac{\delta}{1-\delta}g\mu_B H\Big)^2 + \Big(\frac{\pi}{2}D\Big)^2}}\Big], \quad (S.57)$$

and the absorbed power per unit length is given as $I = \frac{(\pi g \mu_B)^2}{\hbar} H_{\mathrm{rad}}^2 \nu A$.

Finally, the mode intensity ratio (S.55) at a given frequency $\nu$ is given by

$$\mathcal{R} = \frac{I_\nu^{+-}(H_+)}{I_\nu^{+-}(H_-)} = \frac{\sqrt{(\delta h\nu)^2 + (1-\delta^2)^2 (\pi D/2)^2} - \delta h\nu}{\sqrt{(\delta h\nu)^2 + (1-\delta^2)^2 (\pi D/2)^2} + \delta h\nu}. \quad (S.58)$$

## V. RENORMALIZATION GROUP RESULTS

### A. RG theory

Backscattering interaction constant $u$ flows under RG transformation as (remember that $2\delta = u/(2\pi v_F)$)

$$2\delta(\ell) = \frac{2\delta(0)}{1 + 2\delta(0)\ell} \quad (S.59)$$

where $\delta(0) = u(0)/(4\pi v_F) > 0$ is its initial value at the lattice scale and $\ell = \ln(J/E)$ is the logarithmic RG scale, where exchange $J$ serves as the high-energy (lattice scale) cut-off and $E$ is the running energy. Without the magnetic field, for sufficiently low $E$, which means sufficiently large $\ell = \ln(J/E)$ such that $\ell \gg 1/(2\delta(0))$, the first term in the denominator of the above equation (that is, 1) can be neglected and $\delta(\ell)$ reduces to the universal form, independent of the initial value $\delta(0)$,

$$\delta(\ell) \to \frac{1}{2\ell} = \frac{1}{2\ln(J/E)} \to 0 \text{ for } E \to 0. \quad (S.60)$$

That is, it vanishes in the zero energy limit, $E \to 0$. At finite temperature $T$, the minimal value of the energy is set by $T$ and the minimal value of $\delta$ is given by $1/(2\ln(J/T))$. Within the RG terminology, this logarithmically slow vanishing of the interaction $u \propto \delta$ with the energy $E$ and/or temperature $T$ is known as the marginally irrelevant flow. Correspondingly, the interaction amplitude $u(\ell)$ is denoted as the marginally irrelevant one.

In the presence of the field, however, this marginally irrelevant flow is terminated at $E = B$, at which the transverse backscattering $\hat{V}_1$ effectively averages to 0 and the RG flow stops [11]. For energy below this critical value, the coefficient of $\hat{V}_2$ term is given by the finite constant $u_z \approx 4\pi v_F \delta(\ell_b)$, with $\ell_b = \ln(J/B)$, while that of the transverse term $\hat{V}_1$ turns to zero, $u_\perp \to 0$. Importantly, at $E = B$ both coupling constants are finite and given by $u(\ell_b) = 4\pi v_F \delta(\ell_b)$ [11]. This is the essence the Kosterlitz–Thouless RG flow for the spin-1/2 chain in magnetic field, see Fig. 6 in [11] and discussion therein. It is this finite and constant $u = u(\ell_b)$ that shows up in our calculation, simply because we are studying dynamic response of the spin chain at the Zeeman energy $B$, and not at $E \approx 0$ as is often done when one considers the response at the lowest possible energy.

In a remarkable paper [12] Lukyanov has solved exact RG equations for the running coupling $u(T, B)$ as function of both temperature $T$ and magnetic field $B$. According to his Eq. (3.1) his magnetic field $h$ corresponds to our $B$ as $h/2 \to B$. Also, his $2J$ is our $J$, see also Eq. (4.5) in [1]. With these simple changes his key result, Eq. (3.18), can be written in our notations as

$$\frac{1}{2\delta} + \frac{1}{2}\ln(2\delta) = -\mathrm{Re}[\psi(1 + i\frac{B}{2\pi T})] + \ln\Big(\sqrt{\frac{\pi}{2}} e^{1/4}\frac{J}{T}\Big), \quad (S.61)$$

where $\psi(x) = \Gamma'(x)/\Gamma(x)$ is the digamma function and prime denotes derivative with respect to the argument. Asymptotic expansion of digamma function is $\psi(x) \to \ln(x) - 1/(2x)$ for $x \to \infty$. Also, $\psi(1) = -\gamma \approx -0.577$, and $\gamma$ is the Euler constant.

Without the magnetic field, $B = 0$, (S.61) becomes

$$\frac{1}{2\delta} + \frac{1}{2}\ln(2\delta) = \ln\left(\sqrt{\frac{\pi}{2}} e^{\gamma+1/4} \frac{J}{T}\right) \quad (S.62)$$

which is Eq. (4.5) in [1].

At zero temperature, $T \to 0$, we instead use asymptotics of the digamma function to find

$$\frac{1}{2\delta} + \frac{1}{2}\ln(2\delta) = \ln\left(\sqrt{\frac{\pi}{2}} e^{1/4} \frac{2\pi J}{B}\right). \quad (S.63)$$

Remarkably, $T$ under the logarithm cancels out and is replaced by $B$. This is of course just the behavior described qualitatively above — the magnetic field cuts RG flow of the backscattering interaction. Notice that up to the relatively small logarithmic correction due to $\ln(2\delta)$ in the left-hand-side of (S.63), this solution is just $2\delta \approx 1/\ln(\text{cnst} \times J/B)$, just as described above in Eq.(S.60). But the exact solution determines both the constant 'cnst' under the log and the corrections to this simple inverse log behavior.

It is easy to solve (S.61) numerically using Mathematica's built-in digamma function. Doing so it is also easy to find that with the excellent accuracy combined effect of $T$ and $H$ can be accounted for by the simple interpolation $T \to \sqrt{T^2 + e^{2\gamma}(B/(2\pi))^2}$ in (S.62). That is, solving

$$\frac{1}{2\delta} + \frac{1}{2}\ln(2\delta) = \ln\left(\sqrt{\frac{\pi}{2}} e^{\gamma+1/4} \frac{J}{\sqrt{T^2 + \left(\frac{e^\gamma}{2\pi}B\right)^2}}\right) \quad (S.64)$$

one finds $\delta(T, B)$.

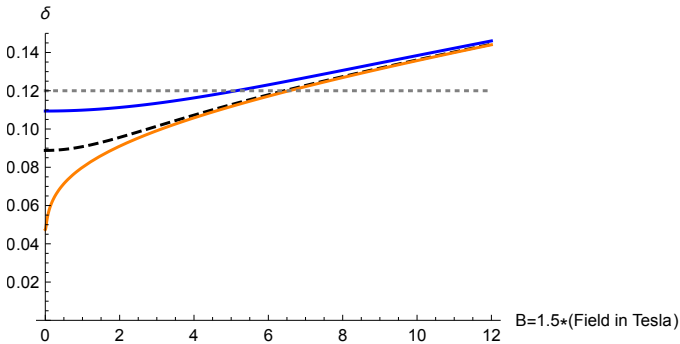

SUPP. FIG. 5. Solution of Eq.(S.61) for $T = 1.3$K (solid blue line), $T = 0.5$K (dashed black line) and $T = 0.0005$K (solid orange line) as a function of field $B$ from 0 to 8 T. The field is converted into Kelvins by $B = 1.5 \times$ (field in Tesla)K. Dotted gray line shows $\delta = 0.12$.

Important lesson of this Figure is that finite temperature $T$ cuts the flow of $\delta$ to zero. For a field of

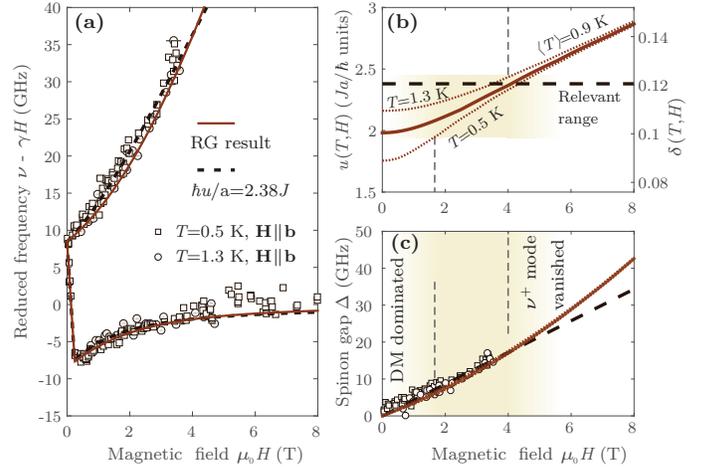

SUPP. FIG. 6. Constant backscattering versus running coupling approach. (a) The "reduced" frequency-field diagram comparing the best fit from Fig. 2b with the exact RG equation (S.64), shown by dashed and solid lines correspondingly. We have removed error bars from the data points here in order to highlight the difference between the two theoretical fits. The RG result gives a good qualitative description, but the systematic deviations from the data are noticeable. (b) The running backscattering interaction parameter $u(H, T)$ at different relevant temperatures (solid and dotted lines) compared to the constant $u$ best-fit value (horizontal dashed line). Shading and the vertical lines delineate the field region that affects the fit the most. The obtained value of $u$ is close to the upper boundary of the RG result. (c) Spinon gap $\Delta$ for the same approaches as in the panel (a) above, and also the experimental value extracted from the spectra ($\Delta = \frac{h}{2}(\nu^+ - \nu^-) - \gamma H$). In practice the impact of temperature turns out to be minor. In addition, this gap is smaller than the magnitude of the DM interaction (and hence $H = 0$ gap) for fields less than 2 T. Above 4 T, the $\nu^+$ mode that is most sensitive to $\Delta$, loses its intensity.

$g\mu_B H = 6$ K, which is $\sim 4$ Tesla, the difference between $\delta$ at $T = 1.3$ K and $T = 0.5$ K is really small.

## B. Comparison with the experiments

The equation (S.64) implies that the positions of the $\nu^\pm$ modes should weakly depend on temperature for a fixed frequency of the spectrometer. However, we do not see much difference between the datasets at 1.3 and 0.5 K. The exact RG result (S.64) with *no* adjustable parameters (but using $\langle T \rangle = 0.9$ K as an 'average' value) is compared with the data and the constant-$u$ fit in Supp. Fig. 6a. While it provides very convincing qualitative description of the data, on a closer look one finds small but systematic deviations for the $\nu^+$ branch. The spinon gap $\Delta$ plotted versus field for the RG approach (Supp. Fig. 6c) shows virtually no change between the relevant temperatures. This is due to the magnetization being very small in the region that is sensitive to the effect of temperature. The corresponding plot of $u(H, T)$ is shown

in Fig. 6b. In addition, the spinon gap $\Delta$ remains small compared to the zero-field gap $\pi D/2$ up to almost 2 T. This means that this region is less sensitive to the details of $u$ behavior. Notice that the small magnitude of $D$ is important for making this field interval smaller. Another insensitive region occurs above 4 T when the $\nu^+$ mode becomes unobservable. This leaves us with a narrow region between 2 and 4 T, where the difference between the RG-based and the constant-$u$ approaches is quite small yet visible.

Both the temperature $T$ and the Zeeman energy $g\mu_B H$ in our experiment are small, but not *very* small compared to $J$. In addition, some other small energy scales unaccounted for in (S.64) may be at play, such as the interchain interaction or the finite-size effects due to the impurities. Eq.(S.80) estimates spinon mean-free path to be about 700 lattice spacings at $T = 1$ K. In principle, given the nominal purity of the starting ingredients being within 99.0 - 99.5% [13], one can naturally expect the typical chain length (and hence the mean free path) being a few hundreds lattice sites. This is already smaller than mean free path ($\propto$ 700 sites) defined by the impurities which introduce a novel characteristic length $\ell_{\rm imp}$. This length is at least partially responsible for the "freezing" of the RG flow at low temperature. This makes unambiguous detection of the RG flow of $u$ impossible. The experimentally found constant best-fit value of $\hbar u/a \simeq 2.38J$ is close to the upper limit of the RG result. This value is also very close to the one derived from the exact diagonalization of the spin-1/2 Heisenberg chain [14] (note that there $\hbar u/a$ is given by $2\pi\lambda_0 = 2.27J$).

We therefore conclude the obtained best-fit value of $u$ represents experimental determination of the *bare*, or the initial, value of the backscattering interaction in the Hamiltonian (S.1). This situation is truly unique and qualitatively different from previous observations of the backscattering interaction in spin-1/2 chains, via low-$T$ behavior of the uniform spin susceptibility [15] and nuclear magnetic relaxation rates measurements in $Sr_2CuO_3$ [16]. In these experimental probes the interaction manifests itself via a weak $\ln(\Lambda/T)$ dependences [10, 17, 18] in which the bare value of $u$ enters only in the under-the-log high-energy cut-off $\Lambda$.

## VI. ADDITIONAL CONSEQUENCES OF THE BACKSCATTERING INTERACTION: SPINON MEAN FREE PATH

It is useful to get some additional intuition about the backscattering interaction $u$. Here we present a calculation of the spinon spin-flip scattering time, following an estimate in [19]. Consider a right-moving spinon with momentum $p$ and spin $s = \uparrow$. Spin-flip part of $u$ (one proportional to $J_R^+ J_L^-$) will cause it to flip a spin to the opposite, spin $\downarrow$. Mathematically, this spin-flip scattering time is given by the imaginary part of the (retarded) self-energy in the diagram in Supp. Fig. 7,

$$\frac{\hbar}{\tau} = 2\,{\rm Im}\Sigma^R(\epsilon = v_F p, p), \quad (S.65)$$

where the self-energy is to be evaluated on the mass-shell, i.e. for $\epsilon = v_F p$.

We start by writing the self-energy in the Matsubara formalism

$$\Sigma(\epsilon_n, p) = -(u/2)^2 T \sum_{\epsilon'_m} \int \frac{dp_1}{2\pi} G(\epsilon'_m, p_1) \Pi(\epsilon_n - \epsilon'_m, p - p_1) \quad (S.66)$$

where $\epsilon$'s are fermion (odd) Matsubara frequencies. We use spectral representation

$$G(\epsilon'_m, p_1) = \int_{-\infty}^{\infty} \frac{dx}{\pi} \frac{{\rm Im}G^R(x, p_1)}{x - i\epsilon'_m} \quad (S.67)$$

and similar for the polarization bubble $\Pi(\epsilon_n - \epsilon'_m, p - p_1)$ of the left-moving spinons. This gives us

$$\Sigma(\epsilon_n, p) = \quad (S.68)$$
$$-(u/2)^2 \int \frac{dp_1}{2\pi} \int_{-\infty}^{\infty} \frac{dx}{\pi} \frac{dy}{\pi} {\rm Im}G^R(x, p_1) {\rm Im}\Pi^R(y, p-p_1) S_1,$$

where

$$S_1 = T \sum_{\epsilon'_m} \frac{1}{x - i\epsilon'_m} \frac{1}{y - i\epsilon_n + i\epsilon'_m}. \quad (S.69)$$

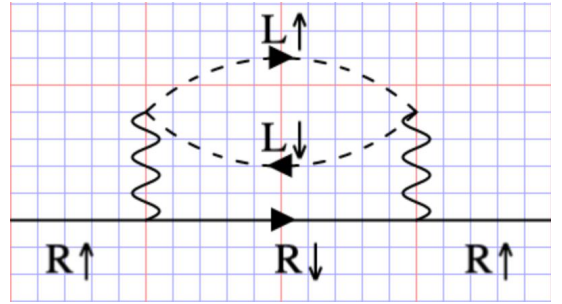

SUPP. FIG. 7. Self-energy diagram describing interaction of the right-moving spinons with the left-moving ones. Wiggly lines represent interaction $u$. Its imaginary part gives spinon scattering rate $1/\tau$.

This sum is evaluated by breaking into elementary fractions and using

$$T\sum_{\epsilon_m} \frac{1}{x - i\epsilon'_m} = T\sum_{m=-\infty}^{\infty} \frac{x}{x^2 + (\pi T(2m+1))^2} = \frac{1}{2}\tanh(x/2T) = \frac{1}{2} - n_F(x), \quad (S.70)$$

where $n_F(x) = 1/(e^{\beta x} + 1)$ is the Fermi distribution function. Next, we see that $\omega_n = \epsilon_n - \epsilon'_m \to 2\pi nT$ is Bose (even) frequency and

$$T\sum_{\omega_n} \frac{1}{y - i\omega_n} = T\sum_{n=-\infty}^{\infty} \frac{y}{y^2 + (2\pi Tn)^2} = \frac{T}{y} + 2T\sum_{n=1}^{\infty} \frac{y}{y^2 + (2\pi Tn)^2} = \frac{1}{2}\coth(y/2T) = \frac{1}{2} + n_B(y), \quad (S.71)$$

where $n_B(y) = 1/(e^{\beta y} - 1)$ is the Bose distribution function.
This leads to

$$\Sigma(\epsilon_n, p) = -(u/2)^2 \int \frac{dp_1}{2\pi} \int_{-\infty}^{\infty} \frac{dx}{\pi} \frac{dy}{\pi} \text{Im} G^R(x, p_1) \text{Im} \Pi^R(y, p - p_1) \frac{1 - n_F(x) + n_B(y)}{x + y - i\epsilon_n}, \quad (S.72)$$

which is easy to continue to the real frequency $i\epsilon_n \to \epsilon + i0$, thereby obtaining the retarded self-energy $\Sigma(\epsilon_n, p) \to \Sigma^R(\epsilon, p)$. We need the imaginary part of that,

$$\text{Im}\Sigma^R(\epsilon, p) = -\frac{(u/2)^2}{\pi} \int_{-\infty}^{\infty} dx dy \, \delta(\epsilon - x - y) \text{Im} G^R(x, p_1) \text{Im} \Pi^R(y, p - p_1) [1 - n_F(x) + n_B(y)]. \quad (S.73)$$

To obtain the imaginary part of the left-spinon bubble $\Pi^R(y, p - p_1)$ we start with its Matsubara expression

$$\Pi(\omega_n, k) = -T\sum_{\epsilon_n} \int \frac{dp_2}{2\pi} G(\epsilon_n, p_2) G(\epsilon_n - \omega_n, p_2 - k) = -T\sum_{\epsilon_n} \int \frac{dp_2}{2\pi} \frac{1}{i\epsilon_n + v_F p_2} \frac{1}{i\epsilon_n - i\omega_n + v_F p_2 - v_F k}. \quad (S.74)$$

This too is broken into elementary fractions and, with the help of (S.70), is brought into a form

$$\Pi(\omega_n, k) = \int \frac{dp_2}{2\pi} \frac{n_F(v_F p_2 - v_F k) - n_F(v_F p_2)}{i\omega_n + v_F k}, \quad (S.75)$$

which is easy to analytically continue $\omega_n \to \omega + i0$ and obtain

$$\text{Im}\Pi^R(\omega, k) = -\pi \int \frac{dp_2}{2\pi} \delta(\omega + v_F k)[n_F(v_F p_2 + \omega) - n_F(v_F p_2)]. \quad (S.76)$$

Plugging (S.76) into (S.73), using $\text{Im} G^R(x, p_1) = -\pi \delta(x - v_F p_1)$ and then integrating over $x$, followed by integration over $y$, using $n_B(-z) = -(1 + n_B(z))$, we obtain

$$\text{Im}\Sigma^R(\epsilon, p) = \pi(u/2)^2 \int \frac{dp_1}{2\pi} \frac{dp_2}{2\pi} \delta(\epsilon + v_F(p - p_1) - v_F p_1)[n_F(v_F p_1) + n_B(v_F(p - p_1))][n_F(v_F p_2 - v_F(p - p_1)) - n_F(v_F p_2)]. \quad (S.77)$$

We see that on the mass-shell $\epsilon = v_F p$ the argument of the delta-function reduces to $2v_F(p - p_1)$, which leads to the divergence of the Bose function $n_B(v_F(p - p_1)) \to T/(v_F(p - p_1))$. But at the same time the last square bracket reduces to the derivative $n_F(v_F p_2 - v_F(p - p_1)) - n_F(v_F p_2) \to \beta v_F(p - p_1) e^{\beta \epsilon}/(e^{\beta \epsilon} + 1)^2$, with $\epsilon = v_F p_2$. Taking this limit carefully we observe that in the $p \to p_1$ limit $n_B(v_F(p-p_1))[n_F(v_F p_2 - v_F(p-p_1)) - n_F(v_F p_2)] \to e^{\beta v_F p_2}/(e^{\beta v_F p_2} + 1)^2$, so that

$$\text{Im}\Sigma^R(v_F p, p) = \frac{(u/2)^2}{4v_F} \int_{-\infty}^{\infty} \frac{dp_2}{2\pi} \frac{e^{\beta v_F p_2}}{(e^{\beta v_F p_2} + 1)^2} = T\frac{u^2}{32\pi v_F^2}. \quad (S.78)$$

Therefore, using $\delta = u/(4\pi v_F)$, we get the final result

$$\frac{\hbar}{\tau} = \pi \delta^2 T. \quad (S.79)$$

Next, the mean-free path of the spin-↑ spinon can be estimated as

$$\ell_{\rm mfp} = v_F\tau = \frac{\hbar v_F}{\pi \delta^2 T} \approx \frac{35 Ja}{T}, \quad (S.80)$$

where $\delta = 0.12$ was used. Thus, for a chain with $J = 20$ K, we have that $\ell_{\rm mfp}/a = 700/T$. The mean-free path is about 700 lattice spacings at $T = 1$ K. This is huge. We see again that even although $u$ is quite large, what really matters is $\delta$, which is quite small.

We note that (S.79) can be also obtained from the simple Fermi's golden rule calculation, similar to how it was done in [19]. The initial state is $|i\rangle = \psi_{R,\uparrow}^\dagger(p)|0\rangle$, while the final state $|f\rangle = \psi_{R,\downarrow}^\dagger(p_1)\psi_{L,\uparrow}^\dagger(p_2+p-p_1)\psi_{L,\downarrow}(p_2)|0\rangle$ involves a particle-hole pair of the left-movers with opposite spins. Using

$$\frac{1}{\tau} = \frac{2\pi}{\hbar}\sum_f |\langle f|\hat{V}_{\rm bs}|i\rangle|^2 \delta(E_f - E_i)\delta(P_f - P_i), \quad (S.81)$$

one recovers result (S.79).

## VII. ESR MEASUREMENTS AND ANALYSIS

### A. $K_2CuSO_4Br_2$ — an ideal model material

The material of our study is bromotinoite $K_2CuSO_4Br_2$, a synthetic analogue of the natural mineral chlorotionite [20]. It provides an outstanding realization of the $S = 1/2$ Heisenberg chain antiferromagnet perturbed by a small uniform DM interaction $\mathbf{D}$ [13, 21, 22]. The presence of the spinon continuum was directly confirmed through the inelastic neutron scattering studies of $K_2CuSO_4Br_2$ [13, 21]. A sketch of the crystal structure is shown in Supp. Fig. 1b. The magnetic $Cu^{2+}$ spin-1/2 ions are forming linear chains running along the **a** axis of the orthorhombic $P_{nma}$ crystal structure. The distance between the magnetic ions is $a = 7.73$ Å. Antiferromagnetic interaction $J \simeq 20.5$ K [21] is mediated by a two-bromine unit, that lacks inversion center within the $ac$ plane. This lack of inversion symmetry naturally gives rise to the DM interaction $\mathbf{D} = (0, 0, \pm D)$ directed along the $z$ axis ($b$ axis of the crystal). A *uniform*, bond-independent arrangement of DM vectors within the chain is a very rare occasion. However, a truly remarkable property of $K_2CuSO_4Br_2$ that distinguishes it from similar materials (e.g. $Cs_2CuCl_4$ [2, 23]) is that the DM axis is the *same*, i.e. oriented along the $b$ crystal axis, for all spin chains [21]. The DM vector is a single-component one, the sign of its $b$ component alternates $\pm D$ between neighboring chains while its two other components are zero. This unique feature makes it possible to apply external magnetic field exactly along the DM vector, which is a crucial element of our study. $K_2CuSO_4Br_2$ turns out to be the material precisely representing the theoretical model of interest.

### B. ESR experiments

The monocrystalline samples of $K_2CuSO_4Br_2$ with typical masses of $10 - 20$ mg were grown at ETH Zürich from aqueous solution according to the techniques described in detiail in Refs. [13, 21]. The crystals are transparent and dark brownish-green in color. They have forms of rectangular or rhombic prisms elongated along the chain $b$ direction. The orientation of the samples was checked with Bruker APEX-II singe crystal diffractometer.

The experiments were done at the Kapitza Institute on a set of multifrequency (1-250 GHz) resonant cavity ESR inserts into $^3$He-pumping cryostat with a 14 T cryomagnet and a $^4$He-pumping cryostat with a 6 T magnet. In either case the sample was mounted into the copper resonator with $\mathbf{H} \parallel \mathbf{b}$ in the location, where the maximum of oscillating magnetic field is expected. The transmission of microwave power through the sample-containing resonator was measured as the function of the magnetic field at a fixed frequency $\nu$. The transmission of the microwave radiation through the sample containing cavity $P$ is affected by the dissipative susceptibility of the spin subsystem and can be approximately expressed as $\Delta P/P \propto \chi''(0, \nu)$, or rather $\chi''(q_{\rm DM}, \nu)$ in the presence of the uniform DM interaction.

### C. Technique basics

Electron spin resonance is a uniquely sensitive probe of the spin dynamics at $q = 0$ and is particularly well suited for probing physics described in this paper, as was convincingly demonstrated previously [2, 24]. Within the linear response theory, the rate of the energy absorption by the sample, which is measured by ESR, is given by the intensity

$$I^{nn}(\nu) = 4\pi V_{\rm sample} H_{\rm rad}^2 \nu \left(-{\rm Im}[\chi^{nn}(q=0,\nu)]\right), \quad (S.82)$$

where $H_{\rm rad}$ is the amplitude of the microwave magnetic field along the direction **n** that the sample is radiated with. The dissipative part of susceptibility $\chi'' = {\rm Im}[\chi]$ here is defined per unit volume with the sign convention following (S.7).

In the frequently employed Faraday geometry $\mathbf{H}_{\rm rad}$ is chosen to be in the plane normal to the static field $\mathbf{H}$. In that case the rate of absorption is controlled by the spin-flip processes and is determined by ${\rm Im}[\chi^{+-}+\chi^{-+}+\chi^{++}+\chi^{--}]/4$ (typically, contributions from $\chi^{\pm\pm}(q=0,\nu)$ are very small, with spectral weight $\propto D^2$). To probe the longitudinal susceptibility ${\rm Im}[\chi^{zz}]$, one needs to use Voigt geometry when $\mathbf{H}_{\rm rad}$ is directed along the external field $\mathbf{H}$.

In the actual experiment the microwave power transmission is measured thorough the resonant cavity, containing the sample. Also, the microwave source signal was amplitude-modulated at frequency of few kHz

(to allow the lock-in detection scheme) and frequency-modulated around $\nu$ (with depth of 100 MHz typically) to avoid detuning effects due to the change in the reactive sample's susceptibility part at the resonance. This second modulation is made in a sawtooth-like mode with a period of about 0.01 s, which is much smaller than the time constant of the lock-in detector. The polarization of the microwave magnetic field is linear, but the actual direction of $\mathbf{H}_{\rm rad}$ in the sample is usually a mixture of Voigt and Faraday configurations. The electrodynamics of such "loaded" cavity is rather complicated [25]. If we define $P_0$ as the power, transmitted with the non-absorbing sample (i.e. for $\chi''$ of interest being zero), the generic formula reads as:

$$P(H) = \frac{P_0}{\left(1 + 2\pi\eta Q \frac{V_{\rm sample}}{V_{\rm resonator}}|\chi''(q=0,\nu,H)|\right)^2}. \quad (S.83)$$

Here $Q$ is the quality factor of the particular cavity mode with eigenfrequency $\nu$, $\eta$ is a geometrical factor of the order of 1. Also the ratio of sample and resonator volumes plays a role. The formula (S.83) essentially describes the interplay between the absorbtion in the sample (proportional to $\nu\chi''$) and the "regular" ohmic losses (also proportional to $\nu$ and field-independent). Thus, in the ESR experiment of this type we directly probe the field dependence of the dissipative part of uniform susceptibility at a given frequency, although without absolute calibration (thanks to a number of frequency- and geometry-related factors). At a frequency corresponding to the resonator mode, amplification of the sensitivity of about $Q \sim 10^3 - 10^4$ may be achieved in comparison with measuring transmission through a waveguide with the same sample.

Even more intuitively, for small sample absorption the transmitted power can be approximated as:

$$P(H) = P_0\left[1 - 4\pi Q\eta\frac{V_{\rm sample}}{V_{\rm resonator}}|\chi''(q=0,\nu,H)|\right], \quad (S.84)$$

Once again, we would like to stress, that in this experiment we straightforwardly probe $\chi''$ and not its derivative with respect ot the field.

In most situations the linear approximation (S.84) is sufficient for the data description. In our case, however, using the exact formula (S.83) may be crucial in order to correctly estimate the ESR line intensities (very weak and very strong within the same record) as we will discuss below.

### D. Fitting the line profiles

A typical ESR line can be conveniently described by a single Lorentian profile:

$$\mathcal{L}(H,\ H_0,\ \Gamma_{1/2},\ \mathcal{A}) = \frac{\mathcal{A}}{1+\left(\dfrac{H-H_0}{\Gamma_{1/2}}\right)^2} \quad (S.85)$$

Here the parameters are the line amplitude $\mathcal{A}$, half-width at half-maximum $\Gamma_{1/2}$, and the resonance field $H_0$. Clearly, while the determination of the resonance field does not require distinguishing between the non-linear and linear approaches (S.83,S.84), for large line amplitudes the former is needed in order to get the resonance integral intensity $I = \pi\mathcal{A}\Gamma_{1/2}$ correctly.

In the present case we deal with multiple overlapping lines. Hence, at each frequency we approximate the quantity of interest

$$2\pi Q\eta\frac{V_{\rm sample}}{V_{\rm resonator}}|\chi''(q=0,\nu,H)| = \quad (S.86)$$
$$\sum_i \mathcal{L}(H,\ H_0^{(i)},\ \Gamma_{1/2}^{(i)},\ \mathcal{A}^{(i)}) + \mathcal{L}(H,\ -H_0^{(i)},\ \Gamma_{1/2}^{(i)},\ \mathcal{A}^{(i)})$$

with a number of independent ESR modes $i$ (up to 3 in our case). An effect of the resonance at the oppositely directed magnetic field (that may be important for broad lines with small resonant fields), which corresponds to a second circular polarized component of a linearly polarized microwave field is explicitly accounted for. Examples of such multi-line fits can be found in Fig. 1a of the main text. Thus, the mode intensities $I^+ = \pi\mathcal{A}^+\Gamma_{1/2}^+$ and $I^- = \pi\mathcal{A}^-\Gamma_{1/2}^-$ that we would determine from the fit, would be proportional to the calculated spectral weights $A^+$, $A^-$ (S.45). Using the *intensity ratio* at the given frequency instead allows us to completely eliminate the uncalibrated prefactors like $Q$ etc from the description. The observed intensity ratio $I^+/I^-$ obtained from the fit can be directly compared to the parameter-free theoretical description (S.55).

---


\fi
\end{document}